\documentstyle[12pt]{article}
\begin{document}

\def\ds{\displaystyle}
\def\beq{\begin{equation}}
\def\eeq{\end{equation}}
\def\bea{\begin{eqnarray}}
\def\eea{\end{eqnarray}}
\def\beeq{\begin{eqnarray}}
\def\eeeq{\end{eqnarray}}
\def\ve{\vert}
\def\vel{\left|}
\def\ver{\right|}
\def\nnb{\nonumber}
\def\ga{\left(}
\def\dr{\right)}
\def\aga{\left\{}
\def\adr{\right\}}
\def\lla{\left<}
\def\rra{\right>}
\def\rar{\rightarrow}
\def\nnb{\nonumber}
\def\la{\langle}
\def\ra{\rangle}
\def\ba{\begin{array}}
\def\ea{\end{array}}
\def\tr{\mbox{Tr}}
\def\ssp{{\Sigma^{*+}}}
\def\sso{{\Sigma^{*0}}}
\def\ssm{{\Sigma^{*-}}}
\def\xis0{{\Xi^{*0}}}
\def\xism{{\Xi^{*-}}}
\def\qs{\la \bar s s \ra}
\def\qu{\la \bar u u \ra}
\def\qd{\la \bar d d \ra}
\def\qq{\la \bar q q \ra}
\def\gGgG{\la g^2 G^2 \ra}
\def\q{\gamma_5 \not\!q}
\def\x{\gamma_5 \not\!x}
\def\g5{\gamma_5}
\def\sb{S_Q^{cf}}
\def\sd{S_d^{be}}
\def\su{S_u^{ad}}
\def\ss{S_s^{??}}
\def\sbp{{S}_Q^{'cf}}
\def\sdp{{S}_d^{'be}}
\def\sup{{S}_u^{'ad}}
\def\ssp{{S}_s^{'??}}
\def\sig{\sigma_{\mu \nu} \gamma_5 p^\mu q^\nu}
\def\fo{f_0(\frac{s_0}{M^2})}
\def\ffi{f_1(\frac{s_0}{M^2})}
\def\fii{f_2(\frac{s_0}{M^2})}
\def\O{{\cal O}}
\def\sl{{\Sigma^0 \Lambda}}
\def\es{\!\!\! &=& \!\!\!}
\def\ap{\!\!\! &\approx& \!\!\!}
\def\ar{&+& \!\!\!}
\def\ek{&-& \!\!\!}
\def\cp{&\times& \!\!\!}
\def\se{\!\!\! &\simeq& \!\!\!}
\def\ses{\!\!\! &\sim& \!\!\!}
\def\kpm{&\pm& \!\!\!}
\def\kmp{&\mp& \!\!\!}


\def\simlt{\stackrel{<}{{}_\sim}}
\def\simgt{\stackrel{>}{{}_\sim}}


\title{
         {\Large
                 {\bf
       Double--lepton polarization asymmetries in the 
   $B \rar K \ell^+ \ell^-$ decay beyond the Standard Model 
                 }
         }
      }

\author{\vspace{1cm}\\
{\small T. M. Aliev \thanks
{e-mail: taliev@metu.edu.tr}\,\,,
V. Bashiry
\,\,,
M. Savc{\i} \thanks
{e-mail: savci@metu.edu.tr}} \\
{\small Physics Department, Middle East Technical University,
06531 Ankara, Turkey} }

\date{}

\begin{titlepage}
\maketitle
\thispagestyle{empty}

\begin{abstract}
General expressions for the double--lepton polarizations in the 
$B \rar K \ell^+ \ell^-$ decay are obtained, using model independent 
effective Hamiltonian, including all possible interactions. Correlations
between the averaged double--lepton polarization asymmetries and the
branching ratio, as well as, the averaged single--lepton polarization 
asymmetry are studied. It is observed that, study of the double--lepton
polarization asymmetries can serve as a good test for establishing new
physics beyond the Standard Model. 
\end{abstract}

~~~PACS numbers: 13.20.He, 12.60.--i, 13.88.+e
\end{titlepage}

\section{Introduction}
Rare B meson decays, induced by flavor--changing neutral current (FCNC)
$b \rar s(d) \ell^+ \ell^-$ transition provide a promising testing ground in
search of the effects beyond the standard model (SM). The FCNC decays which
are forbidden at tree level in the SM, appear at loop level and are very
sensitive to the gauge structure of the SM. Moreover, these decays are also
quite sensitive to the present theories beyond the SM. As is well known,
$B \rar K \ell^+ \ell^-$ and $B \rar K^\ast \ell^+ \ell^-$ decays are
one--loop processes in the SM, governed by the $b \rar s \ell^+ \ell^-$
transition, at quark level. Because of their loop structures, these decays
are suppressed and the relevant branching ratios in the SM are expected to
be of the order of, roughly, $5\times 10^{-7}$ for the $B \rar K \ell^+
\ell^-$ decay, and $1.5\times 10^{-6}$ for the $B \rar K^\ast \ell^+ \ell^-$ 
decay, respectively \cite{R6001}--\cite{R6003}. Recently, Belle \cite{R6004}
and BaBar \cite{R6005} Collaborations announced the following measurements
of the branching ratio for the $B \rar K \ell^+ \ell^-$ decay:
\bea
{\cal B}(B \rar K \ell^+ \ell^-) = \left\{ \begin{array}{lc}
\left( 4.8^{+1.0}_{-0.9} \pm 0.3 \pm 0.1\right) \times
10^{-7}& \cite{R6004}~,\\ \\
\left( 0.65^{+0.14}_{-0.13} \pm 0.04\right) \times
10^{-6}& \cite{R6005}~.\end{array} \right. \nnb
\eea
One of the efficient ways in establishing new physics beyond the SM is the
measurement of the lepton polarization \cite{R6006}--\cite{R6011}.
Polarization of a single--lepton has been studied in the 
$B \rar K^\ast \ell^+ \ell^-$ \cite{R6006}, $B \rar X_s \ell^+ \ell^-$ 
\cite{R6007}--\cite{R6008}, $B \rar K \ell^+ \ell^-$ \cite{R6009}, 
$B \rar \pi(\rho) \ell^+ \ell^-$ \cite{R6010} and $B_s \rar \ell^+ \ell^- 
\gamma$ \cite{R6011} decays. It has been pointed out in \cite{R6012} that,
the study of the polarizations of both leptons provides many additional
observables which can be measured, would be useful in testing the SM and
looking new physics beyond the SM. Polarization asymmetries and
forward--backward asymmetry due to both leptons
have been investigated in the $B \rar X_s \tau^+ \tau^-$ \cite{R6013}, $B \rar
K^\ast \tau^+ \tau^-$ \cite{R6014} and $B \rar K \tau^+ \tau^-$
\cite{R6015} decays in the Minimal Supersymmetric Model, respectively.          

The goal of present work is studying various double--lepton polarizations in
the exclusive $B \rar K \ell^+ \ell^-$ decay using the most general form 
of the effective Hamiltonian, including all possible forms of interactions.
Moreover, we study the correlation between double--lepton polarizations and
single--lepton polarizations. Our purpose in doing so is to find regions in
the new Wilson coefficients parameter space, in which the branching ratio
and single--lepton polarization would agree with the SM prediction, while
double--lepton polarizations would not. Obviously, if such a region does
exist, it is an indication of the fact that the new physics effects can be
established by the measurement only of the double--lepton polarizations.

The paper is organized as follows. In section 2, using the most general
form of the effective Hamiltonian, we obtain the matrix element of the
$B \rar K \ell^+ \ell^-$ decay in terms of form factors relevant to
$B \rar K$ transition and then derive analytical results of double--lepton 
polarization asymmetries. In section 3, we numerically investigate
the correlations of double--lepton asymmetries on branching ratio.
Moreover we analyze the correlation of double--lepton polarization observables
to single--lepton polarizations. This section contains also discussion and
our conclusion.

\section{Double--lepton polarizations}

In this section we calculate the double--lepton polarization asymmetries, using
the most general, model independent form of the effective Hamiltonian.
The effective Hamiltonian for the $b \rar s \ell^+ \ell^-$ transition
in terms of twelve model independent four--Fermi
interactions can be written in the following form:
\bea
\label{e6001}
{\cal H}_{eff} \es \frac{G_F\alpha}{\sqrt{2} \pi}
 V_{ts}V_{tb}^\ast
\Bigg\{ C_{SL} \, \bar s i \sigma_{\mu\nu} \frac{q^\nu}{q^2}\, L \,b
\, \bar \ell \gamma^\mu \ell + C_{BR}\, \bar s i \sigma_{\mu\nu}
\frac{q^\nu}{q^2} \,R\, b \, \bar \ell \gamma^\mu \ell \nnb \\
\ar C_{LL}^{tot}\, \bar s_L \gamma_\mu b_L \,\bar \ell_L \gamma^\mu \ell_L +
C_{LR}^{tot} \,\bar s_L \gamma_\mu b_L \, \bar \ell_R \gamma^\mu \ell_R +
C_{RL} \,\bar s_R \gamma_\mu b_R \,\bar \ell_L \gamma^\mu \ell_L \nnb \\
\ar C_{RR} \,\bar s_R \gamma_\mu b_R \, \bar \ell_R \gamma^\mu \ell_R +
C_{LRLR} \, \bar s_L b_R \,\bar \ell_L \ell_R +
C_{RLLR} \,\bar s_R b_L \,\bar \ell_L \ell_R \nnb \\
\ar C_{LRRL} \,\bar s_L b_R \,\bar \ell_R \ell_L +
C_{RLRL} \,\bar s_R b_L \,\bar \ell_R \ell_L+
C_T\, \bar s \sigma_{\mu\nu} b \,\bar \ell \sigma^{\mu\nu}\ell \nnb \\
\ar i C_{TE}\,\epsilon^{\mu\nu\alpha\beta} \bar s \sigma_{\mu\nu} b \,
\bar \ell \sigma_{\alpha\beta} \ell  \Bigg\}~,
\eea
where $L$ and $R$ in (\ref{e6001}) are
\bea  
L = \frac{1-\gamma_5}{2} ~,~~~~~~ R = \frac{1+\gamma_5}{2}\nnb~,
\eea  
and $C_X$ are the coefficients of the four--Fermi interactions and
$q=p_B-p_K$ is the momentum transfer.
Among twelve Wilson coefficients several already exist in the SM.
For example, the coefficients $C_{SL}$ and $C_{BR}$ in penguin operators
correspond to $-2 m_s C_7^{eff}$ 
and $-2 m_b C_7^{eff}$ in the SM, respectively. The next
four terms in Eq. (\ref{e6001}) are the vector type interactions with
coefficients $C_{LL}^{tot}$, $C_{LR}^{tot}$, $C_{RL}$ and $C_{RR}$. Two of these
vector interactions containing $C_{LL}^{tot}$ and $C_{LR}^{tot}$ do exist in the SM
as well in the form $(C_9^{eff}-C_{10})$ and $(C_9^{eff}+C_{10})$.
Therefore we can say that $C_{LL}^{tot}$ and $C_{LR}^{tot}$ describe the
sum of the contributions from SM and the new physics and they can be written
as
\bea
C_{LL}^{tot} \es C_9^{eff} - C_{10} + C_{LL}~, \nnb \\     
C_{LR}^{tot} \es C_9^{eff} + C_{10} + C_{LR}~, \nnb
\eea
The terms with
coefficients $C_{LRLR}$, $C_{RLLR}$, $C_{LRRL}$ and $C_{RLRL}$ describe
the scalar type interactions. The last two terms with the
coefficients $C_T$ and $C_{TE}$, obviously, describe the tensor type
interactions.    

Exclusive $B \rar K \ell^+ \ell^-$ decay is described by the matrix
element of effective Hamiltonian over $B$ and $K$ meson states, which can be
parametrized in terms of form factors.
It follows from Eq. (\ref{e6001})
that in order to calculate the amplitude of the $B \rar K \ell^+ \ell^-$
decay, the following matrix elements are needed 
\bea
&&\lla K\vel \bar s \gamma_\mu b \ver B \rra~,\nnb \\
&&\lla K \vel \bar s i\sigma_{\mu\nu} q^\nu b \ver B \rra~, \nnb \\
&&\lla K \vel \bar s b \ver B \rra~, \nnb \\
&&\lla K \vel \bar s \sigma_{\mu\nu} b \ver B \rra~. \nnb
\eea

These matrix elements are defined as follows:
\bea
\label{e6002}
\lla K(p_{K}) \vel \bar s \gamma_\mu b \ver B(p_B) \rra =
f_+ \Bigg[ (p_B+p_K)_\mu - \frac{m_B^2-m_K^2}{q^2} \, q_\mu \Bigg] 
+ f_0 \,\frac{m_B^2-m_K^2}{q^2} \, q_\mu~,
\eea
with $f_+(0) = f_0(0)$,
\bea
\label{e6003}
\lla K(p_{K}) \vel \bar s \sigma_{\mu\nu}
 b \ver B(p_B) \rra = -i \, \frac{f_T}{m_B+m_K}
\Big[ (p_B+p_K)_\mu q_\nu -
q_\mu (p_B+p_K)_\nu\Big]~.
\eea

The matrix elements $\lla K(p_{K}) \vel \bar s i \sigma_{\mu\nu} q^\nu b
\ver B(p_B) \rra$ and $\lla K \vel \bar s b \ver B \rra$
can be obtained from Eqs. (\ref{e6002}) and 
(\ref{e6003}). Multiplying both sides of the equations by $q^\mu$, and 
using equation of motion we get
\bea
\label{e6004}
\lla K(p_{K}) \vel \bar s b \ver B(p_B) \rra \es
f_0 \, \frac{m_B^2-m_K^2}{m_b-m_s}~, \\
\label{e6005}                   
\lla K(p_{K}) \vel \bar s i \sigma_{\mu\nu} q^\nu b \ver B(p_B) \rra \es   
\frac{f_T}{m_B+m_K} \Big[ (p_B+p_K)_\mu q^2 -
q_\mu (m_B^2-m_K^2) \Big]~.
\eea
Using the definition of the form factors given in Eqs.
(\ref{e6002})--(\ref{e6004}), we get the amplitude for the 
$B \rar K \ell^+ \ell^-$ decay which can be written as 
\bea
\label{e6006}
{\cal M}(B\rightarrow K \ell^{+}\ell^{-}) \es
\frac{G_F \alpha}{4 \sqrt{2} \pi} V_{tb} V_{ts}^\ast
\Bigg\{
\bar \ell \gamma^\mu \ell \, \Big[
A (p_B+p_K)_\mu + B q_\mu \Big] \nnb \\ 
\ar \bar \ell \gamma^\mu \gamma_5 \ell \, \Big[
C (p_B+p_K)_\mu  + D q_\mu \Big]
+\bar \ell \ell \,Q
+ \bar \ell \gamma_5 \ell \, N \nnb \\
\ar 4 \bar \ell \sigma^{\mu\nu}  \ell\, (- i G ) 
\Big[ (p_B+p_K)_\mu q_\nu - (p_B+p_K)_\nu q_\mu
\Big] \nnb \\
\ar 4 \bar \ell \sigma^{\alpha\beta}  \ell \,
\epsilon_{\mu\nu\alpha\beta} \, H
\Big[ (p_B+p_K)_\mu q_\nu - (p_B+p_K)_\nu q_\mu \Big] 
\Bigg\}~.
\eea

The functions entering to Eq. (\ref{e6004}) are defined as
\bea
\label{e6007}
A \es (C_{LL}^{tot} + C_{LR}^{tot} + C_{RL} + C_{RR})\, f_+ +
2 (C_{BR}+C_{SL}) \,\frac{f_T}{m_B+m_{K}} ~, \nnb \\
B \es (C_{LL}^{tot} + C_{LR}^{tot}+ C_{RL} + C_{RR}) \, f_- -
2 (C_{BR}+C_{SL})\,\frac{f_T}{(m_B+m_{K})q^2}\, (m_B^2-m_K^2) ~, \nnb \\
C \es (C_{LR}^{tot} + C_{RR} - C_{LL}^{tot} - C_{RL})\, f_+ ~,\nnb \\
D \es (C_{LR}^{tot} + C_{RR} - C_{LL}^{tot} - C_{RL})\, f_- ~, \nnb \\
Q \es f_0 \, \frac{m_B^2-m_K^2}{m_b-m_s}\,
(C_{LRLR} + C_{RLLR}+C_{LRRL} + C_{RLRL})~,\nnb \\
N \es f_0 \, \frac{m_B^2-m_K^2}{m_b-m_s}\,
(C_{LRLR} + C_{RLLR}-C_{LRRL} - C_{RLRL})~,\nnb \\
G \es \frac{C_T}{m_B+m_K}\, f_T~,\nnb \\
H \es \frac{C_{TE}}{m_B+m_K}\, f_T~,
\eea
where 
\bea
f_- = (f_0-f_+) \frac{m_B^2-m_K^2}{q^2}~.\nnb
\eea

We see from Eq. (\ref{e6006}) that the difference from the SM is due to the
last four terms only. namely, scalar and tensor type interactions. From the
expression of the matrix element given in Eq. (\ref{e6006}), we get the
following result for the dilepton invariant mass spectrum:
\bea
\label{e6008}
\frac{d\Gamma}{d\hat{s}}(B \rar K \ell^+ \ell^-) = 
\frac{G^2 \alpha^2 m_B}{2^{14} \pi^5}
\vel V_{tb}V_{ts}^\ast \ver^2 \lambda^{1/2}(1,\hat{r}_K,\hat{s}) v
\Delta(\hat{s})~,
\eea
where $\lambda(1,\hat{r}_K,\hat{s})=1+\hat{r}_K^2+\hat{s}^2-2\hat{r}_K-2\hat{s}-
2\hat{r}_K\hat{s}$, $\hat{s}=q^2/m_B^2$, $\hat{r}_K=m_K^2/m_B^2$, 
$\hat{m}_\ell=m_\ell/m_B$, $v=\sqrt{1-4\hat{m}_\ell^2/\hat{s}}$ is the
final lepton velocity, and $\Delta(\hat{s})$ is
\bea  
\label{e6009}
\lefteqn{
\Delta = \frac{4 m_B^2}{3} \mbox{\rm Re}\Big[
-96 \lambda m_B^3 \hat{m}_\ell (A G^\ast) + 
24 m_B^2 \hat{m}_\ell^2 (1-\hat{r}_K) (C D^\ast) + 
12 m_B \hat{m}_\ell (1-\hat{r}_K) (C N^\ast)} \nnb \\
\ar 12 m_B^2 \hat{m}_\ell^2 \hat{s} \,\vel D \ver^2 + 
3 \hat{s} \,\vel N \ver^2 + 12 m_B \hat{m}_\ell \hat{s} (D N^\ast) 
+ 256 \lambda m_B^4 \hat{s} v^2  \,\vel H \ver^2
+ \lambda m_B^2 (3-v^2)\,\vel A \ver^2 \nnb \\
\ar 3 \hat{s} v^2  \,\vel Q \ver^2
+ 64 \lambda m_B^4 \hat{s} (3-2 v^2)\,\vel G \ver^2 
+ m_B^2 \big\{2 \lambda - (1-v^2)\big[ 2 \lambda -
3(1-\hat{r}_K)^2 \big] \big\}\,\vel C \ver^2 \Big]~.
\eea

We now proceed by calculating the double--polarization asymmetries, i.e.,
when polarizations of both leptons are simultaneously measured.
We introduce a spin projection operator defined by
\bea
\Lambda_1 \es \frac{1}{2} (1+\gamma_5\!\!\not\!{s}_i^-)~,\nnb \\
\Lambda_2 \es \frac{1}{2} (1+\gamma_5\!\!\not\!{s}_i^+)~, \nnb
\eea
for lepton $\ell^-$ and antilepton $\ell^+$, where $i=L,N,T$ correspond to
the longitudinal, normal and transversal polarizations, respectively.
Firstly, we define the following orthogonal unit vectors $s^{-\mu}$ in the
rest frame of $\ell^-$ and  $s^{+\mu}$ in the rest frame of $\ell^+$:
\bea
\label{e6010}
s^{-\mu}_L \es \ga 0,\vec{e}_L^{\,-}\dr =
\ga 0,\frac{\vec{p}_-}{\vel\vec{p}_- \ver}\dr~, \nnb \\
s^{-\mu}_N \es \ga 0,\vec{e}_N^{\,-}\dr = \ga 0,\frac{\vec{p}_K\times
\vec{p}_-}{\vel \vec{p}_K\times \vec{p}_- \ver}\dr~, \nnb \\
s^{-\mu}_T \es \ga 0,\vec{e}_T^{\,-}\dr = \ga 0,\vec{e}_N^{\,-}
\times \vec{e}_L^{\,-} \dr~, \nnb \\
s^{+\mu}_L \es \ga 0,\vec{e}_L^{\,+}\dr =
\ga 0,\frac{\vec{p}_+}{\vel\vec{p}_+ \ver}\dr~, \nnb \\
s^{+\mu}_N \es \ga 0,\vec{e}_N^{\,+}\dr = \ga 0,\frac{\vec{p}_K\times
\vec{p}_+}{\vel \vec{p}_K\times \vec{p}_+ \ver}\dr~, \nnb \\
s^{+\mu}_T \es \ga 0,\vec{e}_T^{\,+}\dr = \ga 0,\vec{e}_N^{\,+}
\times \vec{e}_L^{\,+}\dr~,
\eea
where $\vec{p}_\mp$ and $\vec{p}_K$ are the three--momenta of the
leptons $\ell^\mp$ and K meson in the
center of mass frame (CM) of $\ell^- \,\ell^+$ system, respectively.

The longitudinal unit vectors $s^-_L$ and $s^+_L$ are boosted to CM frame of
the $\ell^- \,\ell^+$ system by the Lorentz transformation, giving
\bea
\label{e6011}
\ga s^{-\mu}_L \dr_{CM} \es \ga \frac{\vel\vec{p}_- \ver}{m_\ell}~,
\frac{E \vec{p}_-}{m_\ell \vel\vec{p}_- \ver}\dr~,\nnb \\
\ga s^{+\mu}_L \dr_{CM} \es \ga \frac{\vel\vec{p}_- \ver}{m_\ell}~,
-\frac{E \vec{p}_-}{m_\ell \vel\vec{p}_- \ver}\dr~,
\eea
while the vectors $s^{\mp\mu}_N$ and $s^{\mp\mu}_T$ are not changed by the boost.

We can now define the double--lepton polarization asymmetries as in
\cite{R6012}:
\bea
\label{e6012}
P_{ij}(\hat{s}) =
\frac{\ds{\Bigg( \frac{d\Gamma}{d\hat{s}}(\vec{s}_i^-,\vec{s}_j^+)}-
\ds{\frac{d\Gamma}{d\hat{s}}(-\vec{s}_i^-,\vec{s}_j^+) \Bigg)} -
\ds{\Bigg( \frac{d\Gamma}{d\hat{s}}(\vec{s}_i^-,-\vec{s}_j^+)} -
\ds{\frac{d\Gamma}{d\hat{s}}(-\vec{s}_i^-,-\vec{s}_j^+)\Bigg)}}
{\ds{\Bigg( \frac{d\Gamma}{d\hat{s}}(\vec{s}_i^-,\vec{s}_j^+)} +
\ds{\frac{d\Gamma}{d\hat{s}}(-\vec{s}_i^-,\vec{s}_j^+) \Bigg)} +      
\ds{\Bigg( \frac{d\Gamma}{d\hat{s}}(\vec{s}_i^-,-\vec{s}_j^+)} +
\ds{\frac{d\Gamma}{d\hat{s}}(-\vec{s}_i^-,-\vec{s}_j^+)\Bigg)}}~,
\eea
where $i,j=L,~N,~T$, and the first subindex $i$ corresponds 
lepton while the second subindex $j$ corresponds to antilepton,
respectively.

After lengthy calculations we get the following results for the 
double--polarization asymmetries.

\bea
\label{e6013}
P_{LL} \es \frac{4 m_B^2}{3\Delta} \, \mbox{\rm Re} \Big[
32 \lambda m_B^3 \hat{m}_\ell (A^\ast G) +
24 m_B^2 \hat{m}_\ell^2 (1-\hat{r}_K) (C^\ast D) \nnb \\
\ar 12 m_B \hat{m}_\ell (1-\hat{r}_K) (C^\ast N)
+ 256 \lambda m_B^4 \hat{s} v^2 \vel H \ver^2 -
64 \lambda m_B^4 \hat{s} (1- 2 v^2) \vel G \ver^2 \nnb \\
\ek \lambda m_B^2 (1+v^2) \vel A \ver^2
+ 12 m_B^2 \hat{m}_\ell^2 \hat{s}\vel D \ver^2 + 
3 \hat{s} \vel N \ver^2 + 12 m_B \hat{m}_\ell \hat{s} (D^\ast N)
+ 3 \hat{s} v^2 \vel Q \ver^2 \nnb \\
\ek m_B^2 \big\{ 2 \lambda -(1-v^2) \big[2 \lambda + 
3 (1-\hat{r}_K)^2 \big] \big\} \vel C \ver^2 \Big]~,
\\ \nnb \\
\label{e6014}
P_{LN} \es  \frac{2\pi m_B^3 \sqrt{\lambda
\hat{s}}}{\hat{s}\Delta}
\, \mbox{\rm Im} \Big[
2 m_B \hat{m}_\ell \hat{s} \mbox{\rm Im}(A^\ast D) +
32 m_B^2 \hat{m}_\ell^2 \hat{s} (D^\ast G) +
\hat{s} (A^\ast N) \nnb \\
\ek 16 m_B \hat{m}_\ell \hat{s} (G^\ast N)
- \hat{s} v^2 (C^\ast Q)
+ 2 m_B \hat{m}_\ell (1-\hat{r}_K) (A^\ast C) \nnb \\
\ar 32 m_B^2 \hat{m}_\ell^2 (1-\hat{r}_K) (C^\ast G)
\Big]~, \\ \nnb \\
\label{e6015}
P_{NL} \es \frac{2\pi m_B^3 \sqrt{\lambda
\hat{s}}}{\hat{s}\Delta}
\, \mbox{\rm Im} \Big[
- 2 m_B \hat{m}_\ell \hat{s} (A^\ast D) 
- 32 m_B^2 \hat{m}_\ell^2 \hat{s} (D^\ast G)
- \hat{s} (A^\ast N) \nnb \\
\ar 16 m_B \hat{m}_\ell \hat{s} (G^\ast N)
- \hat{s} v^2 (C^\ast Q)
- 2 m_B \hat{m}_\ell (1-\hat{r}_K) (A^\ast C) \nnb \\
\ek 32 m_B^2 \hat{m}_\ell^2 (1-\hat{r}_K) (C^\ast G)
\Big]~, \\ \nnb \\ \nnb   
\label{e6016}
P_{LT} \es \frac{2\pi m_B^3 \sqrt{\lambda
\hat{s}}}{\hat{s}\Delta}
\, \mbox{\rm Re} \Big[
2 m_B \hat{m}_\ell (1-\hat{r}_K) v \vel C \ver^2 
+ 2 m_B \hat{m}_\ell \hat{s} v (C^\ast D)
+ \hat{s} v (C^\ast N) \nnb \\
\ek \hat{s} v (A^\ast Q)
+ 16 m_B \hat{m}_\ell \hat{s} v (G^\ast Q)
\Big]~, \\ \nnb \\
\label{e6017}
P_{TL} \es \frac{2\pi m_B^3 \sqrt{\lambda
\hat{s}}}{\hat{s}\Delta}
\, \mbox{\rm Re} \Big[
2 m_B \hat{m}_\ell (1-\hat{r}_K) v \vel C \ver^2 
+ 2 m_B \hat{m}_\ell \hat{s} v (C^\ast D)
+ \hat{s} v (C^\ast N) \nnb \\
\ar \hat{s} v (A^\ast Q)
- 16 m_B \hat{m}_\ell \hat{s} v (G^\ast Q)
\Big]~, \\ \nnb \\
\label{e6018}
P_{NT} \es \frac{8 m_B^2 v}{3\Delta} 
\, \mbox{\rm Im} \Big[
- 32 \lambda m_B^3 \hat{m}_\ell (A^\ast H)
+ 128  \lambda m_B^4 \hat{s} (G^\ast H)
+ 6 m_B \hat{m}_\ell \hat{s} (D^\ast Q) \nnb \\
\ar 3 \hat{s} (N^\ast Q)
- 2 \lambda m_B^2 (A^\ast C)
- 32 \lambda m_B^3 \hat{m}_\ell (C^\ast G)
+ 6 m_B \hat{m}_\ell (1-\hat{r}_K) (C^\ast Q)
\Big]~, \\ \nnb \\ 
\label{e6019}
P_{TN} \es \frac{8 m_B^2 v}{3\Delta} 
\, \mbox{\rm Im} \Big[
- 32 \lambda m_B^3 \hat{m}_\ell (A^\ast H)
+ 128  \lambda m_B^4 \hat{s} (G^\ast H)
+ 6 m_B \hat{m}_\ell \hat{s} (D^\ast Q) \nnb \\
\ar 3 \hat{s} (N^\ast Q)
+ 2 \lambda m_B^2 (A^\ast C)
+ 32 \lambda m_B^3 \hat{m}_\ell (C^\ast G)
+ 6 m_B \hat{m}_\ell (1-\hat{r}_K) (C^\ast Q)
\Big]~, \\ \nnb \\
\label{e6020}
P_{TT} \es \frac{4 m_B^2}{3\Delta} 
\, \mbox{\rm Re} \Big[
32 \lambda m_B^3 \hat{m}_\ell (A^\ast G)
- 24 m_B^2 \hat{m}_\ell^2 (1-\hat{r}_K) (C^\ast D)
- 12 m_B \hat{m}_\ell (1-\hat{r}_K) (C^\ast N) \nnb \\
\ek 256 \lambda m_B^4 \hat{s} v^2 \vel H \ver^2
- 64 \lambda m_B^4 \hat{s} (1-2 v^2) \vel G \ver^2
- \lambda m_B^2 (1+v^2) \vel A \ver^2 \nnb \\
\ek 12 m_B^2 \hat{m}_\ell^2 \hat{s} \vel D \ver^2 
- 3 \hat{s} \vel N \ver^2
-12 m_B \hat{m}_\ell \hat{s} (D^\ast N)
+ 3 \hat{s} v^2 \vel Q \ver^2 \nnb \\
\ar m_B^2 \big\{ 2 \lambda - (1-v^2) \big[2 \lambda +
3 (1-\hat{r}_K)^2 \big] \big\} \vel C \ver^2  
\Big]~, \\ \nnb \\
\label{e6021}
P_{NN} \es \frac{4 m_B^2}{3\Delta}
\, \mbox{\rm Re} \Big[
96 \lambda m_B^3 \hat{m}_\ell (A^\ast G)
+ 256 \lambda m_B^4 \hat{s} v^2 \vel H \ver^2
- 3 \hat{s} v^2 \vel Q \ver^2
+ 12 m_B^2 \hat{m}_\ell^2 \hat{s} \vel D \ver^2 \nnb \\ 
\ar 3 \hat{s} \vel N \ver^2
+ 12 m_B \hat{m}_\ell \hat{s} (D^\ast N)
- \lambda m_B^2 (3-v^2) \vel A \ver^2
- 64 \lambda m_B^4 \hat{s} (3-2 v^2) \vel G \ver^2 \nnb \\ 
\ar m_B^2 \big\{ 2 \lambda - (1-v^2) \big[2 \lambda -
3 (1-\hat{r}_K)^2 \big] \big\} \vel C \ver^2
+ 24 m_B^2 \hat{m}_\ell^2 (1-\hat{r}_K) (C^\ast D) \nnb \\
\ar 12 m_B \hat{m}_\ell (1-\hat{r}_K) (C^\ast N)
\Big]~. 
\eea

\section{Numerical results and discussion}
In this section we present the numerical analysis of all 
possible double--lepton polarizations, whose explicit expressions we give in
the previous section.

The values of the input parameters used in this work are:
$\vel V_{tb} V_{ts}^\ast\ver = 0.0385$, $(C_9^{eff})^{sh}= 4.344$,
$C_{10}=-4.669$, $\Gamma_B = 4.22\times 10^{-13}~GeV$. It is well known that
the Wilson coefficient $C_9^{eff}$ receives long distance contribution
coming from the real intermediate $J/\psi$ family. However, in the present
work we consider only the short distance contribution. The modulo of
$C_7^{eff}$ is fixed by the experimental value of ${\cal B}(B \rar X_s
\gamma)$, while its sign is determined by the SM. In further analysis we
use $(C_7^{eff})_{SM}=-0.313$ and for the parametrization of the form factors 
we use the results of the first reference in \cite{R6003}.

The region for the new Wilson coefficients can be obtained from
existing experimental results of BaBar and BELLE Collaboration on   
${\cal B}(B \rar K \ell^- \ell^+)$ \cite{R6004,R6005} (see figures below). 

It follows from 
Eqs. (\ref{e6013})--(\ref{e6021}) that double--lepton polarization asymmetries 
depend on $q^2$ and
the new Wilson coefficients. Therefore, it may experimentally be difficult
to study these dependencies at the same time. For this reason, we eliminate
$q^2$ dependence by performing integration over $q^2$ in the allowed region,
i.e., we consider the averaged double--lepton polarization asymmetries. The
averaging over $q^2$ is defined as
\bea
\la P_{ij} \ra = \frac{\ds \int_{4 \hat{m}_\ell^2}^{(1-\sqrt{\hat{r}_K})^2} 
P_{ij} \frac{d{\cal B}}{d \hat{s}} d \hat{s}}
{\ds \int_{4 \hat{m}_\ell^2}^{(1-\sqrt{\hat{r}_K})^2} 
\frac{d{\cal B}}{d \hat{s}} d \hat{s}}~.\nnb
\eea   
We present our analysis in a series of figures. In Figs. (1)--(4), we depict 
the correlation of the averaged double--lepton asymmetries on the branching 
ratio for the $B \rar K \mu^- \mu^+$ decay. Note that the region of the
branching ratio is taken from the existing experimental result, and the
corresponding regions of variation of the new Wilson coefficients are given
in the figures. 

From these figures we deduce the following results:

\begin{itemize}

\item{ 
There exist regions of new Wilson coefficients where 
$\la P_{LL} \ra$ departs the SM result considerably when 
${\cal B}(B \rar K \mu^- \mu^+)$ is very close to SM value.
     }

\item{
$\la P_{LN}\ra$, as well as $\la P_{NL}\ra$, seem to exceed the SM 
value 3--4 times, and they change their signs when new Wilson coefficients 
vary in the allowed region and branching ratio is very close to the SM 
result. This behavior can serve as a good test for establishing new physics 
beyond the SM.
     }

\item{
In the presence of the new Wilson coefficients, the value of 
$\la P_{LT}\ra$ ($\la P_{TL}\ra$) is 3--4 times
smaller(larger) compared to the SM prediction. Moreover, $\la P_{TL}\ra$
changes its sign when new Wilson coefficients vary.
     }

\end{itemize} 

We do not not present the correlation of $\la P_{NN}\ra$, $\la P_{NT}\ra$,
$\la P_{TN}\ra$ and $\la P_{TT}\ra$ on the branching ratio, since the
values of $\la P_{NN}\ra$, $\la P_{NT}\ra$ and $\la P_{TN}\ra$ are
very small, and the behavior of $\la P_{TT}\ra$ is quite similar to that
of $\la P_{TL}\ra$. Change in the values of $\la P_{NT}\ra$ and 
$\la P_{TN}\ra$ is observed, but no change in their signs seems to occur.
 
In Figs. (5)--(13), we present the correlation of $\la P_{ij} \ra$ 
on branching ratio for the $B \rar K \tau^+ \tau^-$ decay. Similar to the
$B \rar K \mu^+ \mu^-$ decay, one concludes that several $\la P_{ij} \ra$
are sizable and sensitive to the existence of new physics. It  should be
noted that, in the present analysis we change the branching ratio in the
region $(1 \div 3.5)\times 10^{-7}$.

Next, we want to discuss the following problem. Can we establish the new
physics effects only by measuring the double--lepton polarization. In other
words, do sizable regions of new Wilson coefficients exist, for which the
single--lepton polarization coincides with the SM result, while
double--lepton polarizations do not. In order to analyze this possibility,
we study the correlations of averaged double $\la P_{ij} \ra$ and
single--lepton $\la P_{i} \ra$ polarizations. We vary the new Wilson
coefficients in the region allowed by the measured branching ratio.

Our numerical analysis shows that, for the $B \rar K \mu^+ \mu^-$ case, the
correlations $(\la P_{TL} \ra,~\la P_{L} \ra)$ and
$(\la P_{LT} \ra,~\la P_{T} \ra)$ are more informative. The correlations 
$(\la P_{LL} \ra,~\la P_{L} \ra)$ and
$(\la P_{TT} \ra,~\la P_{T} \ra)$ are not suitable since their values in
the SM are practically the same and if the new Wilson coefficients are taken
into account in the allowed region, the departure of $\la P_{LL} \ra$ and
$\la P_{TT} \ra$ from their SM values is very small. In Figs. (14) and
(15) we present the correlations of $\la P_{TL} \ra$ on $\la P_{L} \ra$
and $\la P_{LT} \ra$ on $\la P_{T} \ra$, respectively. From these
figures we observe that, there exist regions of the new Wilson coefficients,
where double--lepton polarizations differ from the SM, while single--lepton
polarizations coincide with the SM prediction. Here in this figure and in
rest of the  following ones, the numbers in the parentheses are the values 
of the branching ratio corresponding to the respective lower and upper 
values of the new Wilson coefficients.  

The situation for the $B \rar K \tau^+ \tau^-$ decay is slightly different.
We obtain that the study of all correlations between
double-- and single--lepton polarizations leads to strong restriction on
tensor type Wilson coefficient $C_T$. Besides, analyses of the correlations 
$(\la P_{TT} \ra,~\la P_{T} \ra)$, $(\la P_{LT} \ra,~\la P_{T} \ra)$
and $(\la P_{TL} \ra,~\la P_{T} \ra)$ show that there exist regions of
the new Wilson coefficients $C_{RR},~C_{LR}$ and scalar type coefficients
$C_{LRRL},~C_{RLRL}$ where double--lepton polarizations differ from the SM
results, but single--lepton polarizations coincide with that of the SM
(see Figs. (16), (17) and (18)).

Finally, let us briefly discuss the problem of detectability of the lepton
polarization asymmetries in experiments. Experimentally, to measure an
asymmetry $\la P_{ij} \ra$ of the decay with the branching ratio ${\cal B}$
at $n \sigma$ level, the required relevant number of events 
(i.e., the number of $B \bar{B}$ pair) are given by the expression
\bea
N = \frac{n^2}{{\cal B} s_1 s_2 \la P_{ij} \ra^2}~,\nnb
\eea
where $s_1$ and $s_2$ are the efficiencies of the leptons. Typical values of
the efficiencies of the $\tau$--leptons range from $50\%$ to $90\%$ for their
various decay modes (see for example \cite{R6016} and references therein). 
It should be noted here that the error in $\tau$--lepton 
polarization is estimated to be about $(10 \div 15)\%$ \cite{R6017}.
So, the error in measurement of the $\tau$--lepton asymmetries is of the
order of $(20 \div 30)\%$, and the error in obtaining the number of events is
about $50\%$.

It follows from the expression for $N$ that, in order to observe the lepton
polarization asymmetries in $B \rar K \mu^+ \mu^-$ and $B \rar K \tau^+
\tau^-$ decays at $3\sigma$ level, the minimum number of required events
are (for the efficiency of $\tau$--lepton we take $0.5$):

\begin{itemize}
\item for $B \rar K \mu^+ \mu^-$ decay
\bea
N = \left\{ \begin{array}{ll}
3.5 \times 10^{7}  & (\mbox{\rm for} \lla P_{LL} \rra,\lla P_{LT} \rra)~,\\
5.0 \times 10^{8}  & (\mbox{\rm for} \lla P_{TL} \rra)~,\\
2.0 \times 10^{11} & (\mbox{\rm for} \lla P_{LN} \rra)~,\end{array} \right. \nnb
\eea

\item for $B \rar K \tau^+ \tau^-$ decay
\bea
N = \left\{ \begin{array}{ll}
(1.0 \pm 0.5) \times 10^{9}  & (\mbox{\rm for} \lla P_{LL} \rra,\lla P_{LT}
\rra,\lla P_{TL} \rra,\lla P_{NN} \rra)~,\\   
(5.0 \pm 2.5) \times 10^{8}  & (\mbox{\rm for} \lla P_{TT} \rra)~,\\     
(4.0 \pm 2.0) \times 10^{10} & (\mbox{\rm for} \lla P_{LN} \rra,
\lla P_{NL} \rra)~,\\
(3.0 \pm 1.5) \times 10^{11} & (\mbox{\rm for} \lla P_{NT} \rra,
\lla P_{TN} \rra)~.\end{array} \right.
\nnb
\eea
\end{itemize}

On the other hand, the number of $B \bar{B}$ pairs, that are produced
at B--factories and LHC \, are about $\sim 5\times 10^8$ and $10^{12}$,
respectively. As a result of a comparison of these numbers and $N$, we
conclude that, except $\lla P_{LN} \rra$ in the $B \rar K \mu^+ \mu^-$ decay
and  $\lla P_{NT} \rra$, $\lla P_{TN} \rra$  in the 
$B \rar K \tau^+ \tau^-$ decay,
all double lepton polarizations can definitely be detectable at LHC. The
numbers for the $B \rar K \mu^+ \mu^-$ decay presented above demonstrate
that, $\lla P_{LL} \rra$ and $\lla P_{LT} \rra$ for the $B \rar K \mu^+
\mu^-$ decay should be accessible at $B$ factories after several years of
running.   

In summary, in this work, we present the most general analysis of the
double--lepton polarization asymmetries in the $B \rar K \ell^+ \ell^-$
decay using the most general, model independent form of the effective
Hamiltonian. In our analysis we have used the experimental result of the
branching ratio for the $B \rar K \mu^+ \mu^-$ decay, announced by the BaBar
and BELLE Collaborations. The correlation of the averaged double--lepton
polarization asymmetries on the branching ratio (we use the experimental
result for the varying  region of the branching ratio for the 
$B \rar K \mu^+ \mu^-$ decay). We find out that the study of double--lepton
polarization asymmetries can serve as good test for establishing
new physics beyond the SM. Moreover, we study the correlations between
double-- and single--lepton polarization asymmetries and observe that there
exist regions of the new Wilson coefficients for which double--lepton
polarization asymmetries depart considerably from the SM, while
single--lepton polarization coincides with that of the SM predictions. In
other words, in these regions of the new Wilson coefficients only
double--lepton polarization asymmetry measurements can establish new physics
beyond the SM.

\newpage

\section*{Figure captions}
{\bf Fig. (1)} Parametric plot of the correlation between the averaged 
double--lepton polarization asymmetry $\la P_{LL} \ra$ 
and the branching ratio for the $B \rar K \mu^+ \mu^-$ decay, when 
both leptons are longitudinally polarized.\\ \\
{\bf Fig. (2)} The same as in Fig. (1), but for the averaged
double--lepton polarization asymmetry $\la P_{LN} \ra$,
when one the leptons is longitudinally, and the other is normally
polarized.\\ \\
{\bf Fig. (3)} The same as in Fig. (2), but for the averaged
double--lepton polarization asymmetry $\la P_{LT} \ra$.\\ \\
{\bf Fig. (4)} The same as in Fig. (2), but for the averaged
double--lepton polarization asymmetry $\la P_{TL} \ra$.\\ \\
{\bf Fig. (5)} The same as in Fig. (1), but for the 
$B \rar K \tau^+ \tau^-$ decay.\\ \\
{\bf Fig. (6)} The same as in Fig. (2), but for the 
$B \rar K \tau^+ \tau^-$ decay.\\ \\
{\bf Fig. (7)} The same as in Fig. (5), but for the averaged
double--lepton polarization asymmetry $\la P_{NL} \ra$.\\ \\
{\bf Fig. (8)} The same as in Fig. (3), but for the 
$B \rar K \tau^+ \tau^-$ decay.\\ \\
{\bf Fig. (9)} The same as in Fig. (4), but for the 
$B \rar K \tau^+ \tau^-$ decay.\\ \\
{\bf Fig. (10)} The same as in Fig. (5), but for the averaged
double--lepton polarization asymmetry $\la P_{NN} \ra$.\\ \\
{\bf Fig. (11)} The same as in Fig. (5), but for the averaged
double--lepton polarization asymmetry $\la P_{NT} \ra$.\\ \\
{\bf Fig. (12)} The same as in Fig. (5), but for the averaged
double--lepton polarization asymmetry $\la P_{TN} \ra$.\\ \\
{\bf Fig. (13)} The same as in Fig. (5), but for the averaged
double--lepton polarization asymmetry $\la P_{TT} \ra$,
when both leptons are transversally polarized.\\ \\
{\bf Fig. (14)} Parametric plot of the correlation between the averaged
double--lepton polarization asymmetry $\la P_{TL} \ra$ 
and the single--lepton polarization $\la P_{L} \ra$ for the $B \rar K
\mu^+ \mu^-$ decay. The numbers in the parentheses are the values
of the branching ratio corresponding to the respective lower and upper
values of the new Wilson coefficients. \\ \\
{\bf Fig. (15)} The same as in Fig. (14), but the correlation between
$\la P_{LT} \ra$ and $\la P_{T} \ra$ pair.\\ \\
{\bf Fig. (16)} The same as in Fig. (14), but the correlation between 
$\la P_{TT} \ra$ and $\la P_{T} \ra$ pair, for the
$B \rar K \tau^+ \tau^-$ decay.\\ \\
{\bf Fig. (17)} The same as in Fig. (16), but the correlation between 
$\la P_{LT} \ra$ and $\la P_{T} \ra$ pair.\\ \\
{\bf Fig. (18)} The same as in Fig. (16), but the correlation between 
$\la P_{TL} \ra$ and $\la P_{T} \ra$ pair.

\newpage

\begin{figure}
\vskip 1.5 cm
    \includegraphics{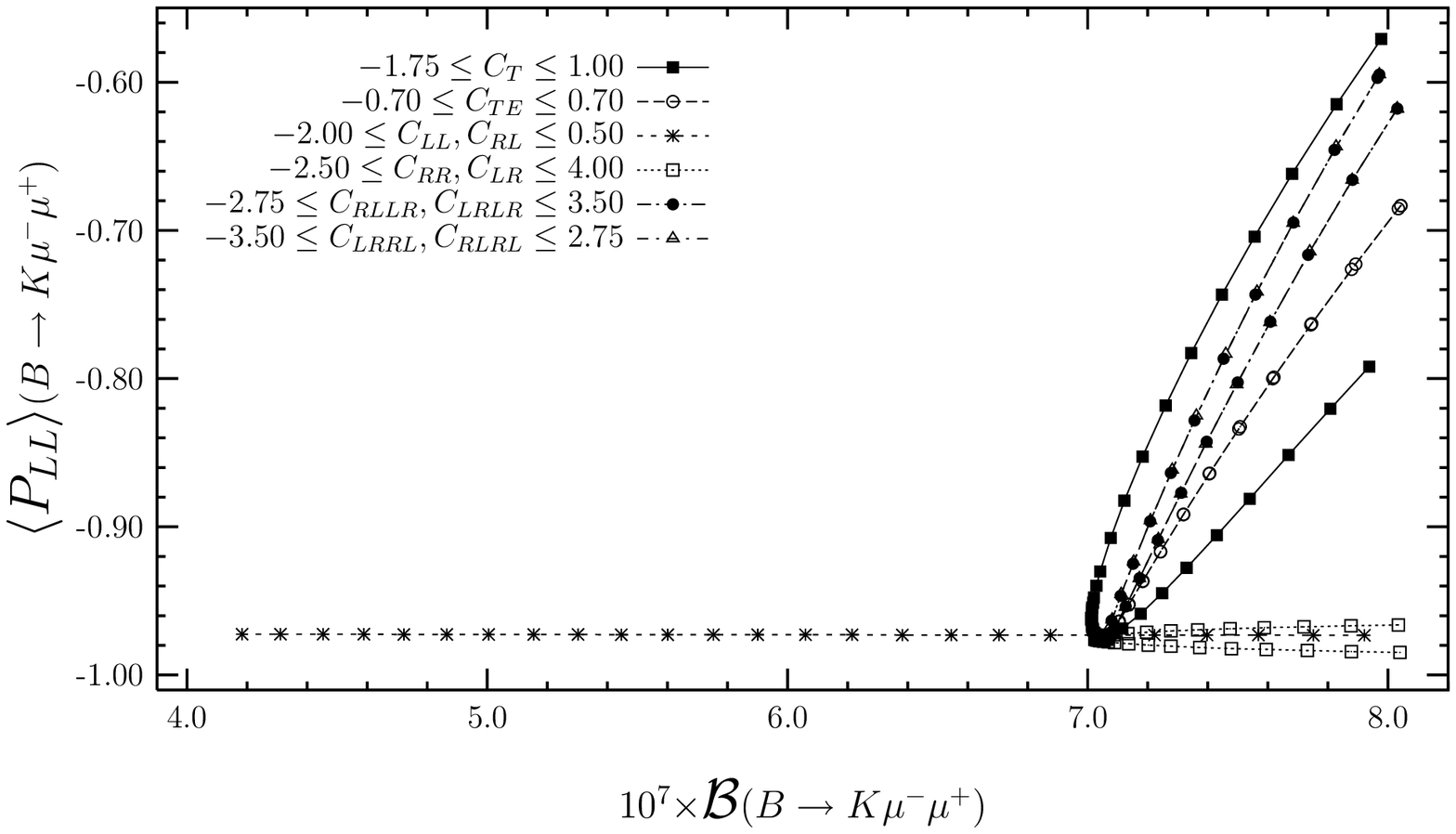}
\vskip 7.8cm
\caption{}
\end{figure}

\begin{figure}
\vskip 2.5 cm
    \includegraphics{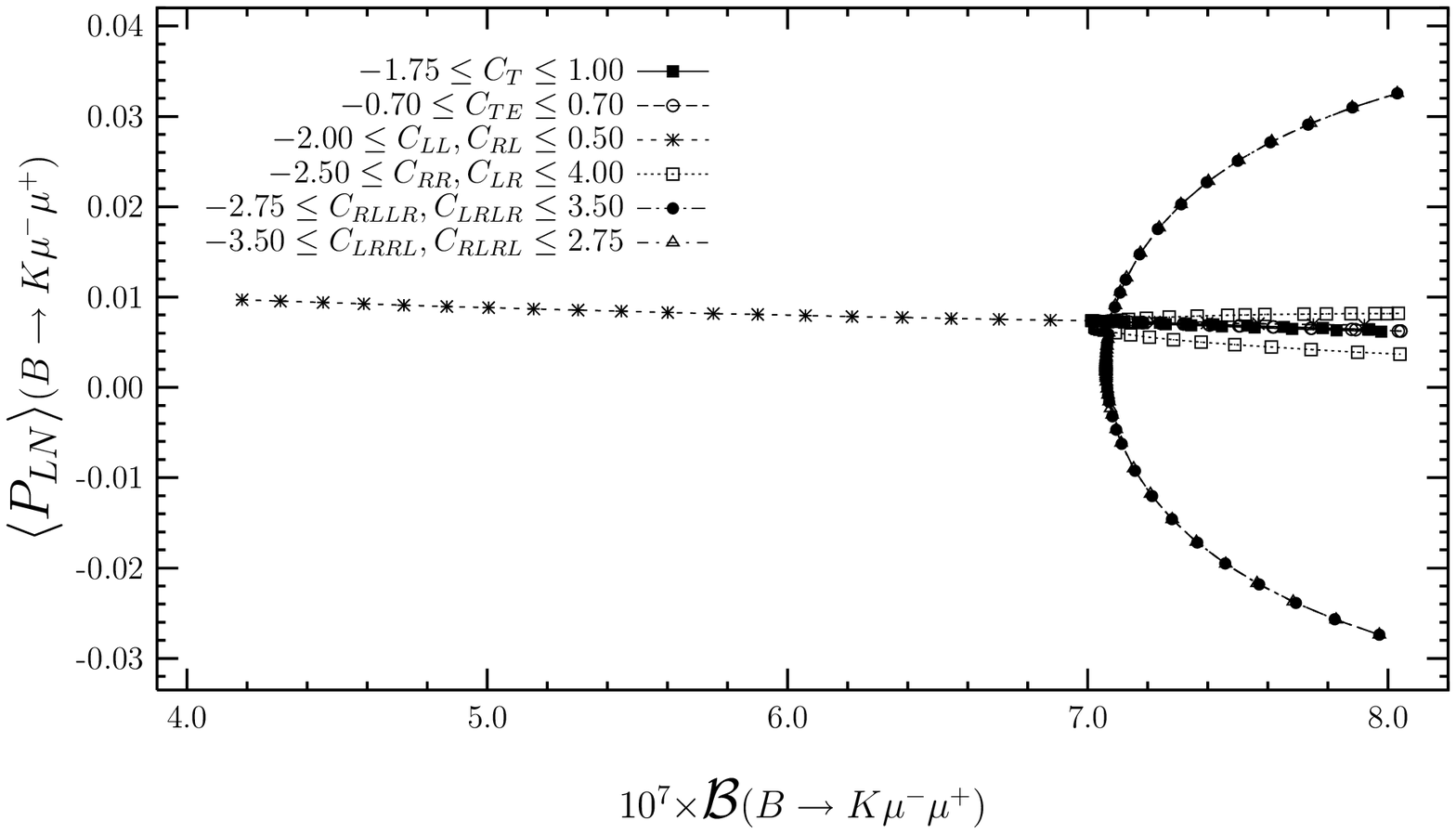}
\vskip 7.8 cm
\caption{}
\end{figure}

\begin{figure}
\vskip 1.5 cm
    \includegraphics{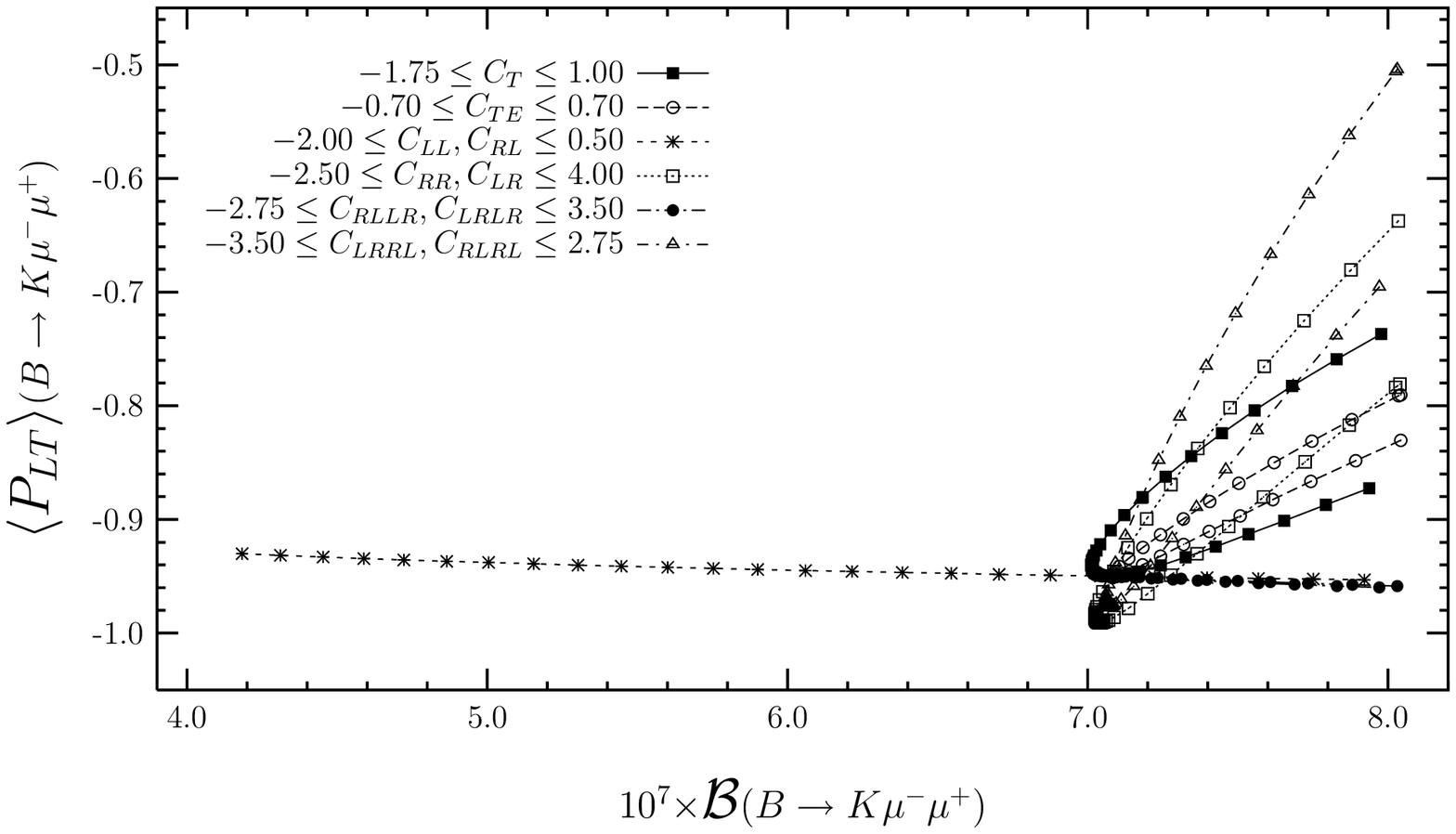}
\vskip 7.8cm
\caption{}
\end{figure}

\begin{figure}
\vskip 2.5 cm
    \includegraphics{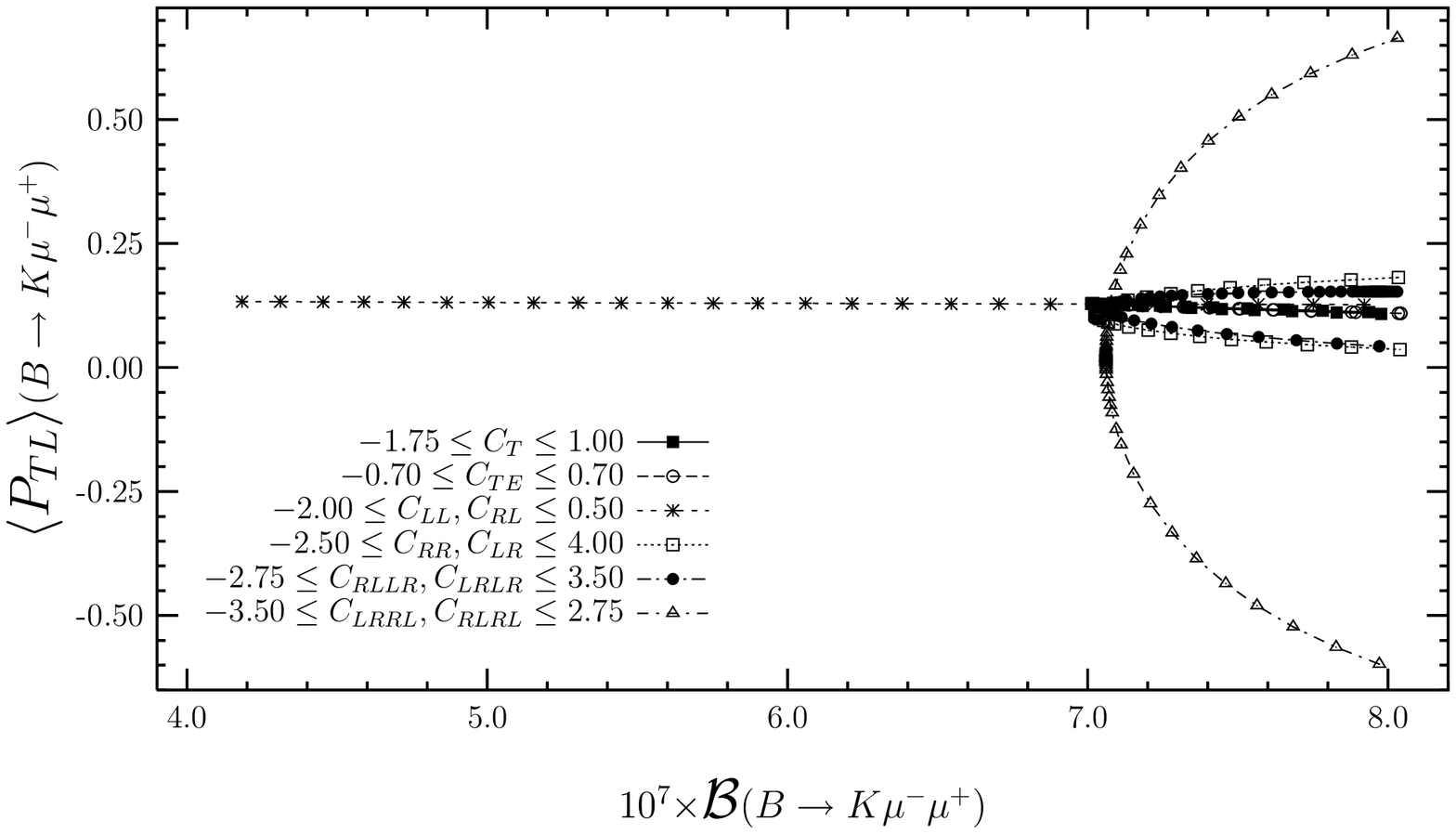}
\vskip 7.8 cm
\caption{}
\end{figure}

\begin{figure}
\vskip 2.5 cm
    \includegraphics{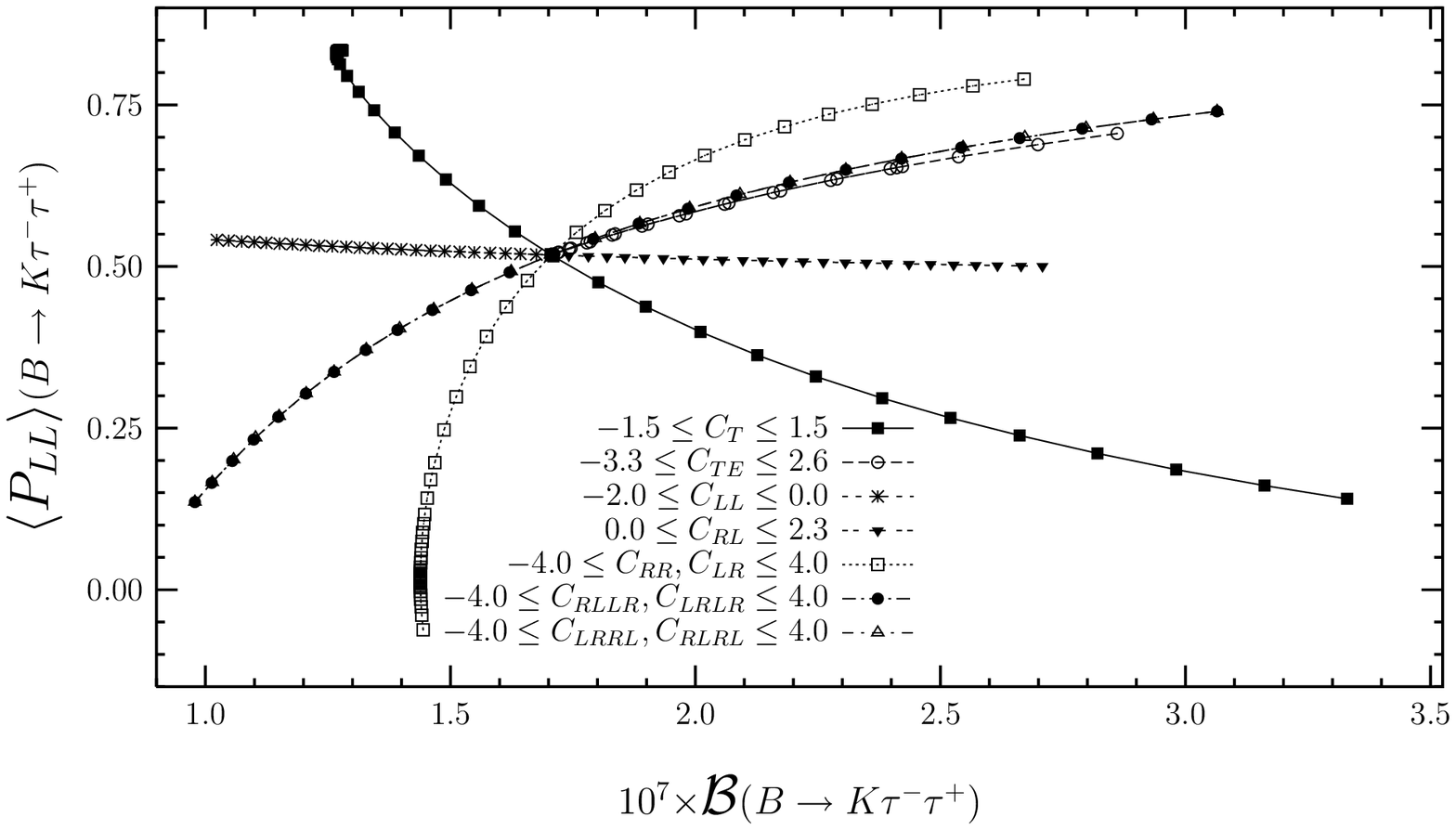}
\vskip 7.8 cm
\caption{}
\end{figure}

\begin{figure}
\vskip 1.5 cm
    \includegraphics{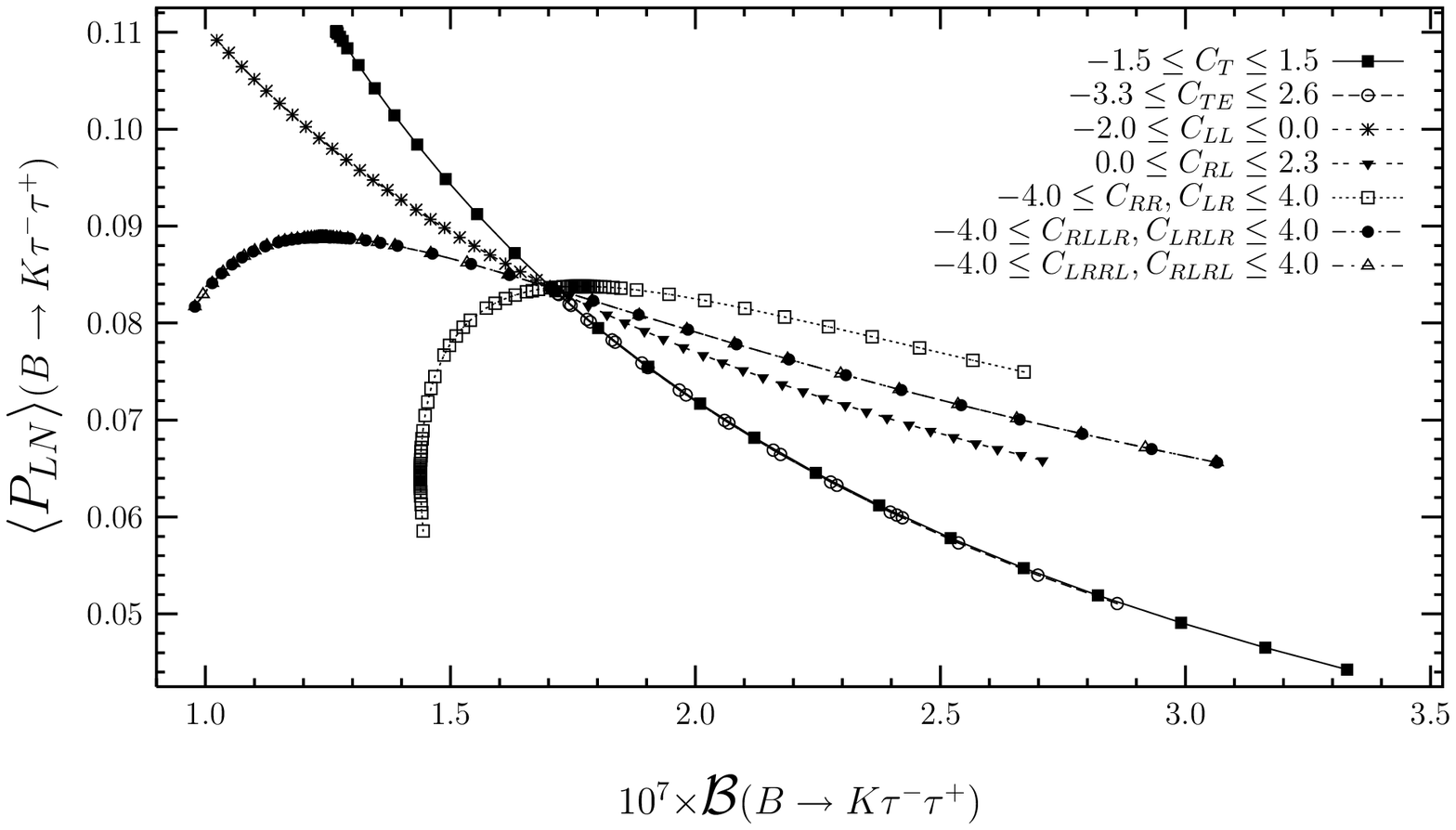}
\vskip 7.8cm
\caption{}
\end{figure}

\begin{figure}
\vskip 2.5 cm
    \includegraphics{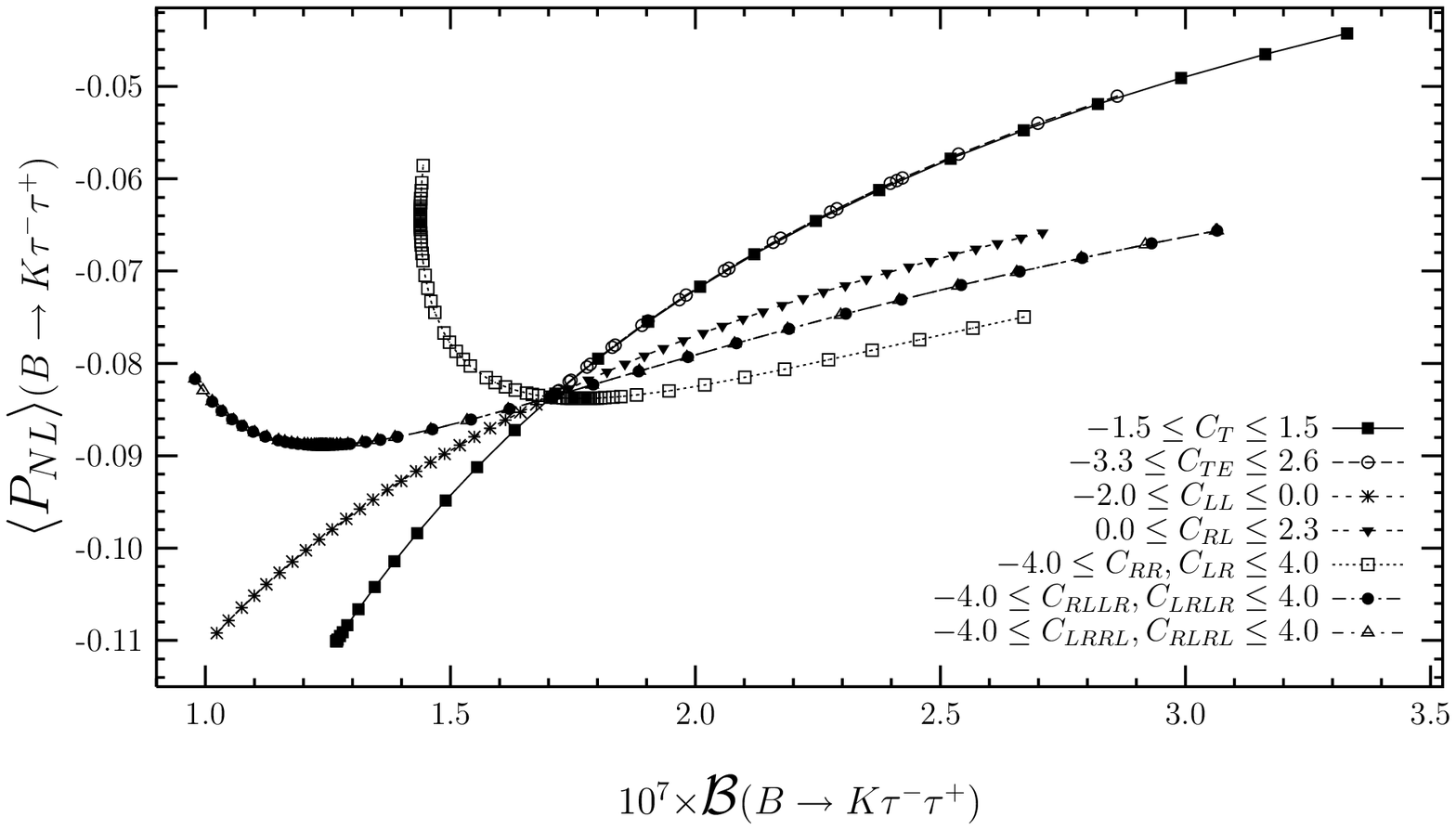}
\vskip 7.8 cm
\caption{}
\end{figure}

\begin{figure}
\vskip 1.5 cm
    \includegraphics{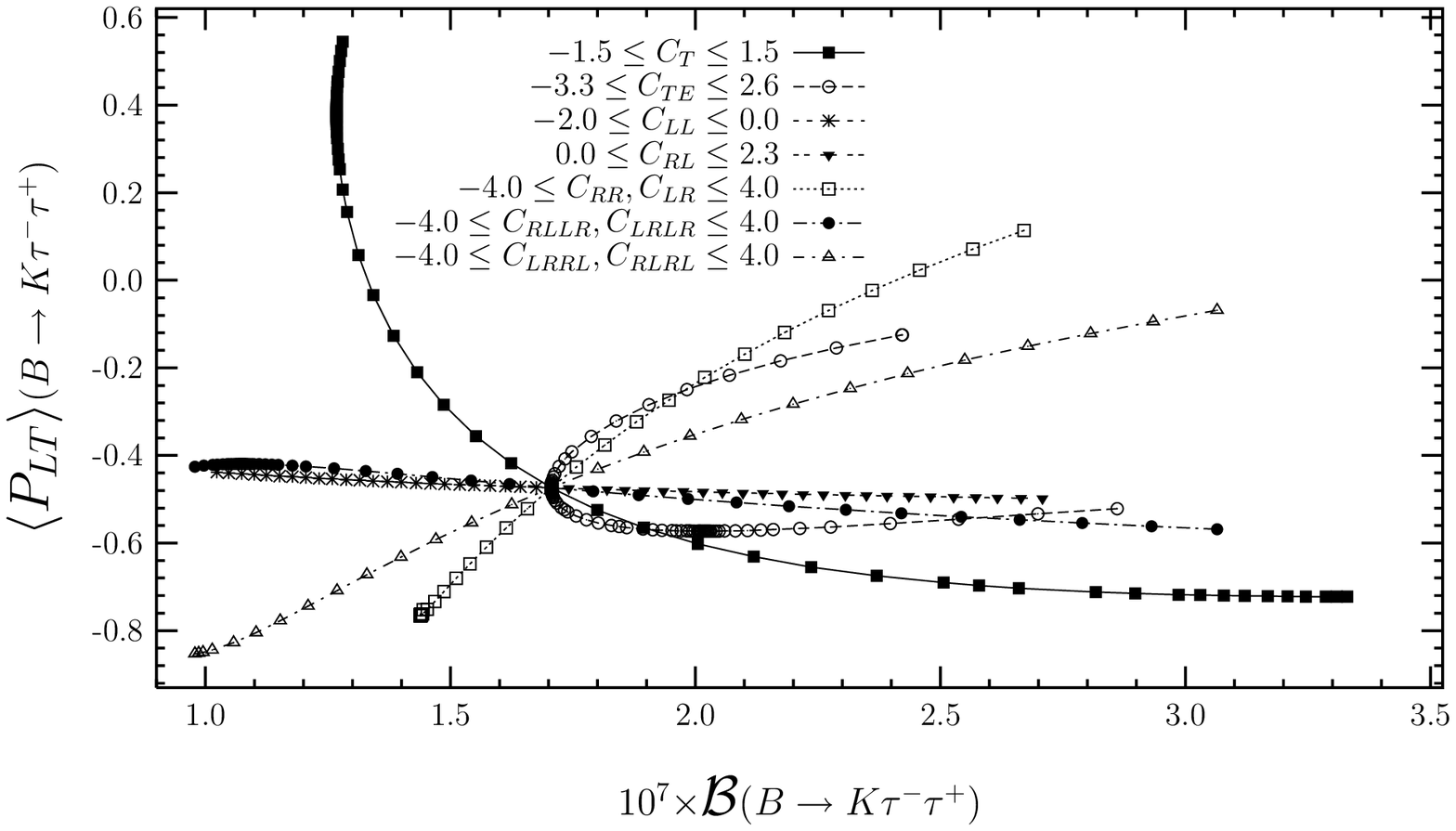}
\vskip 7.8cm
\caption{}
\end{figure}

\begin{figure}
\vskip 2.5 cm
    \includegraphics{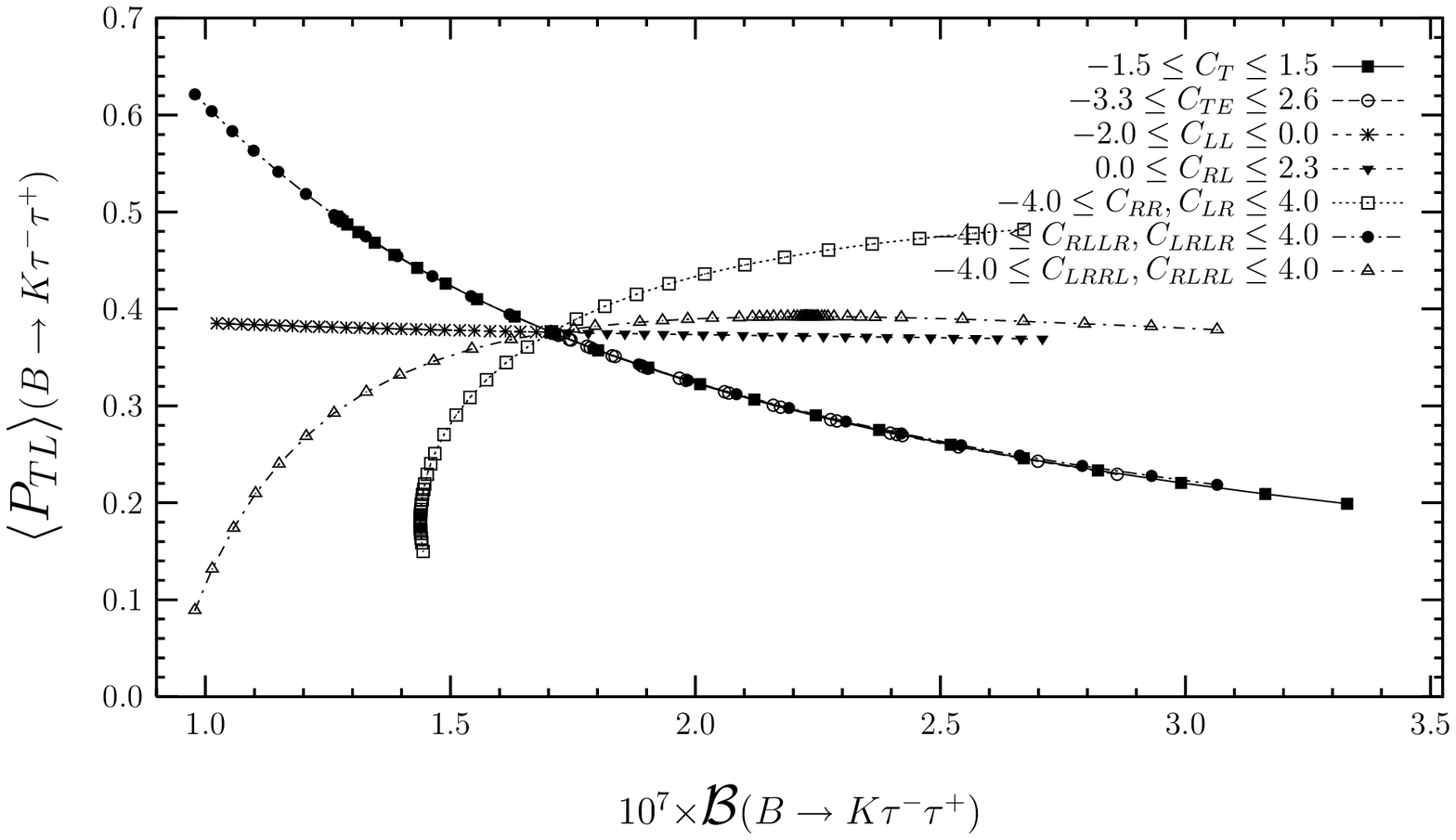}
\vskip 7.8 cm
\caption{}
\end{figure}

\begin{figure}
\vskip 1.5 cm
    \includegraphics{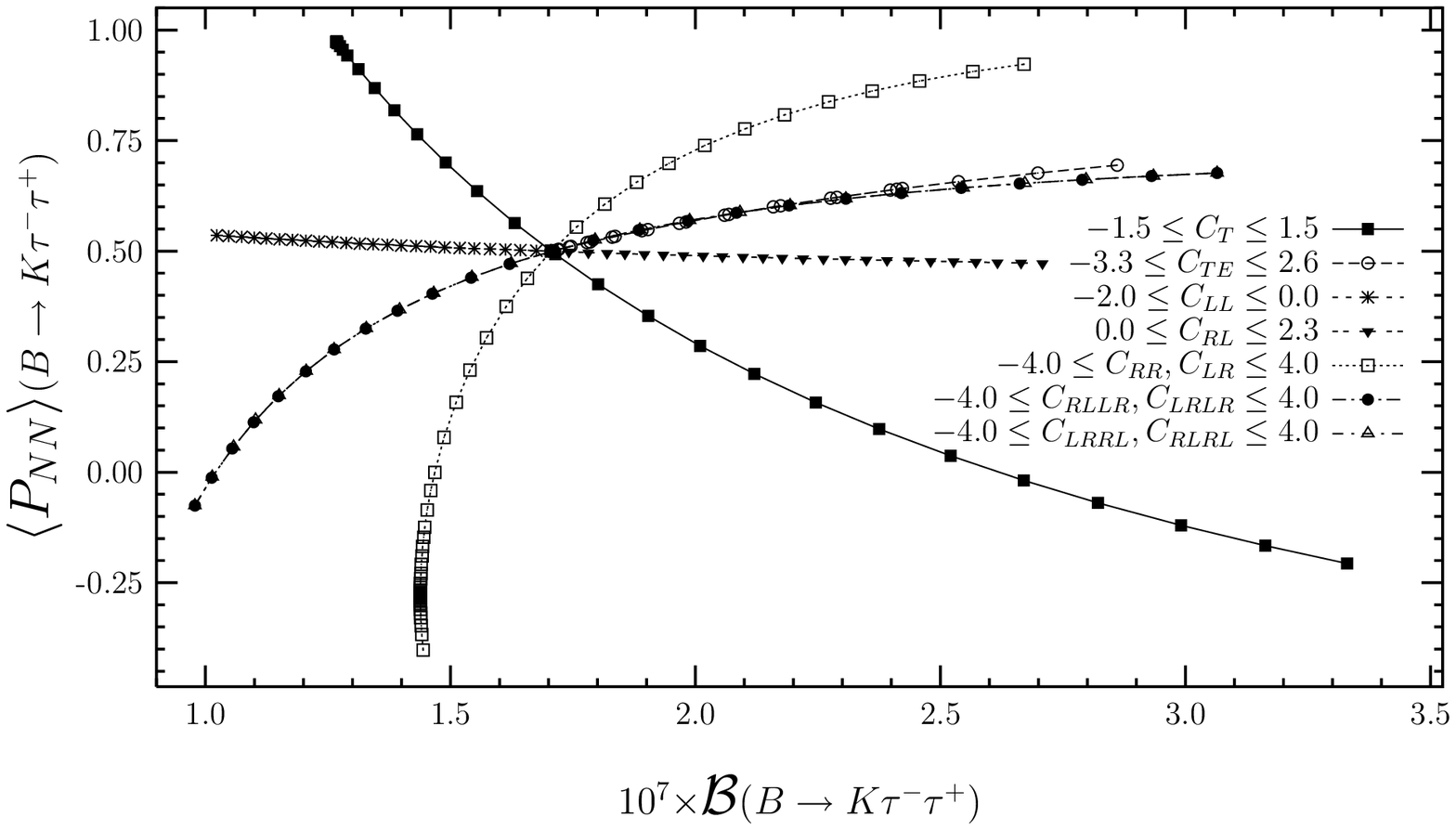}
\vskip 7.8cm
\caption{}
\end{figure}

\begin{figure}
\vskip 2.5 cm
    \includegraphics{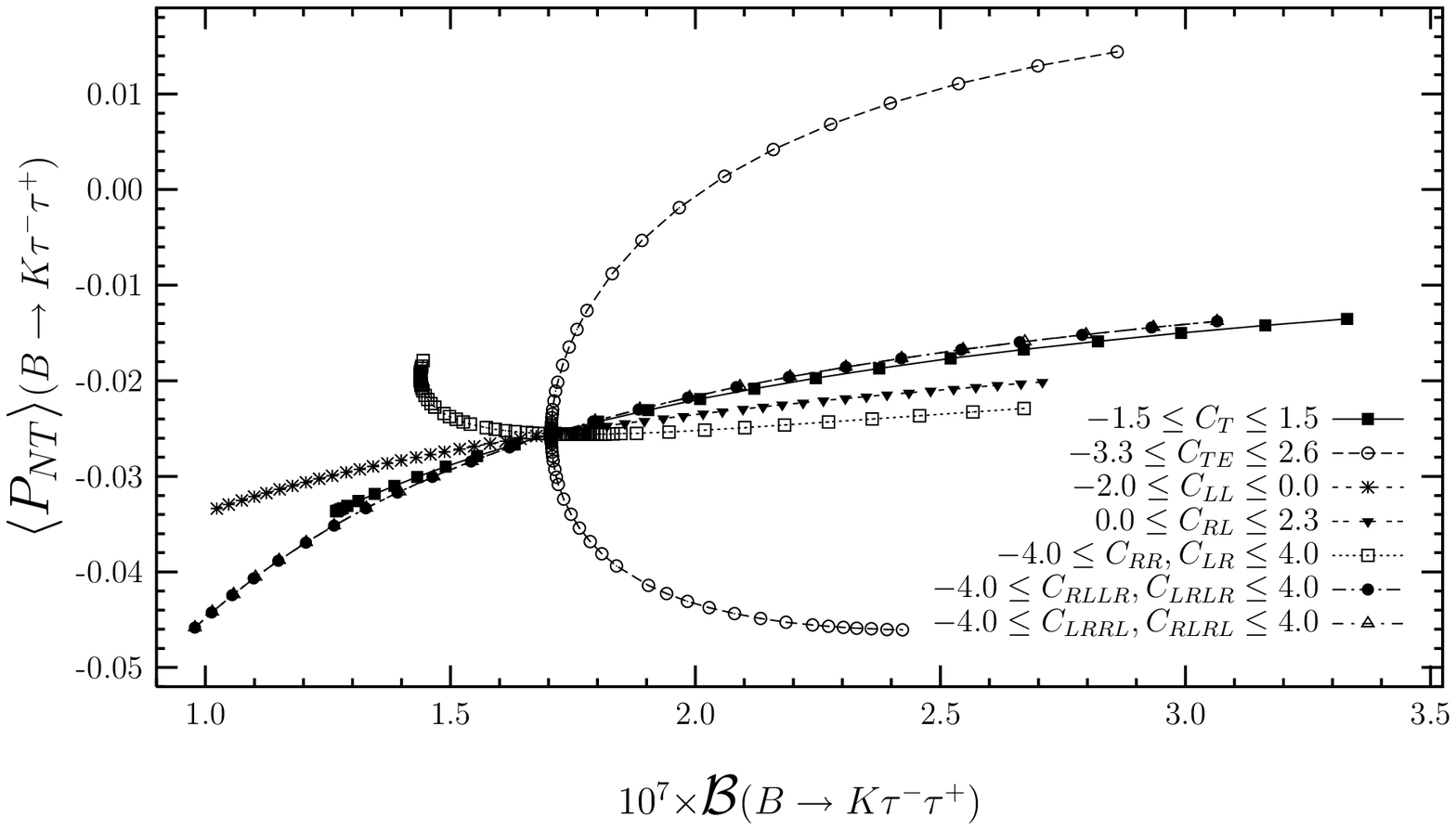}
\vskip 7.8 cm
\caption{}
\end{figure}

\begin{figure}
\vskip 1.5 cm
    \includegraphics{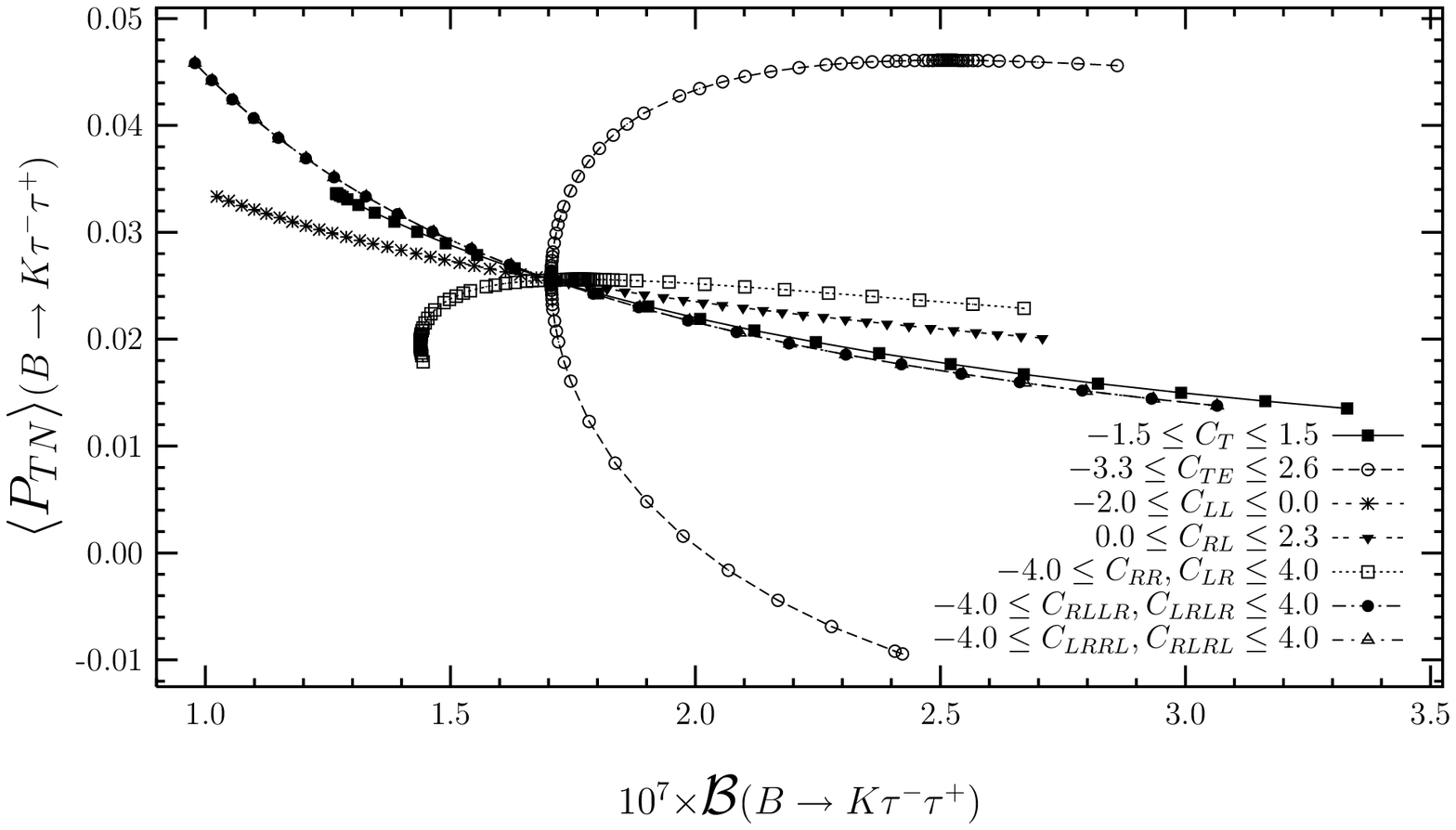}
\vskip 7.8cm
\caption{}
\end{figure}

\begin{figure}
\vskip 2.5 cm
    \includegraphics{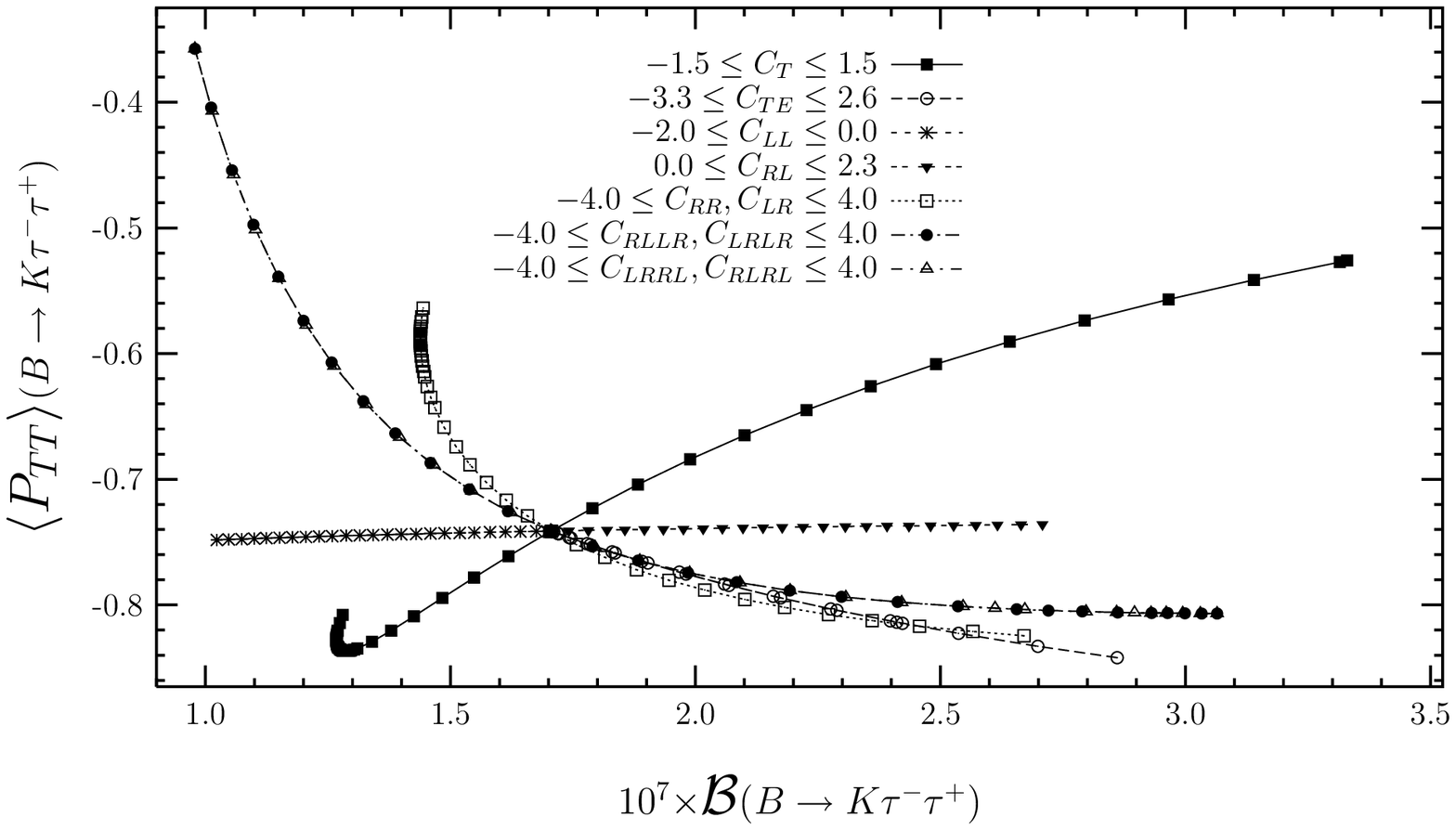}
\vskip 7.8 cm
\caption{}
\end{figure}

\begin{figure}
\vskip 1.5 cm
    \includegraphics{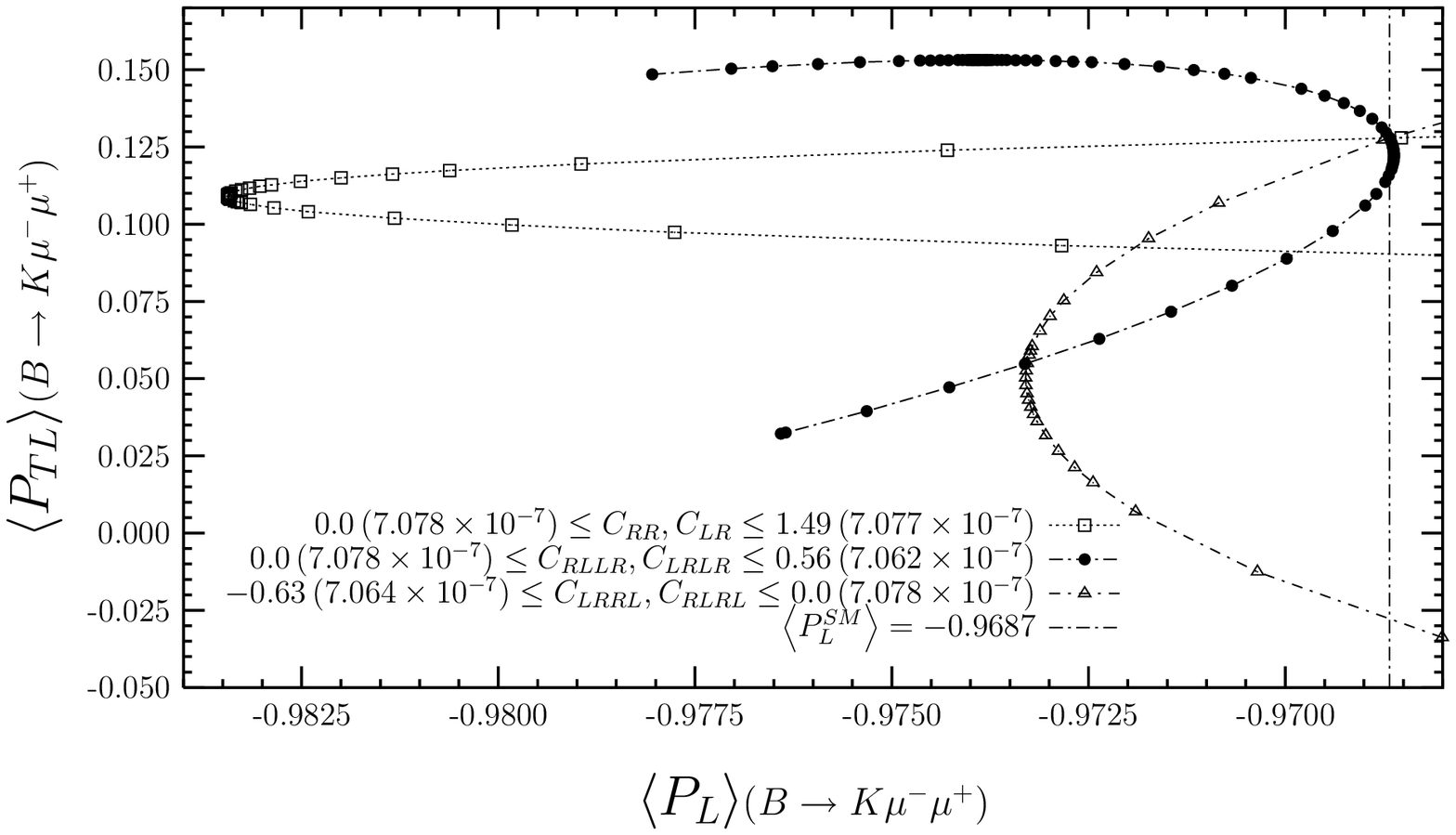}
\vskip 7.8cm
\caption{}
\end{figure}

\begin{figure}
\vskip 2.5 cm
    \includegraphics{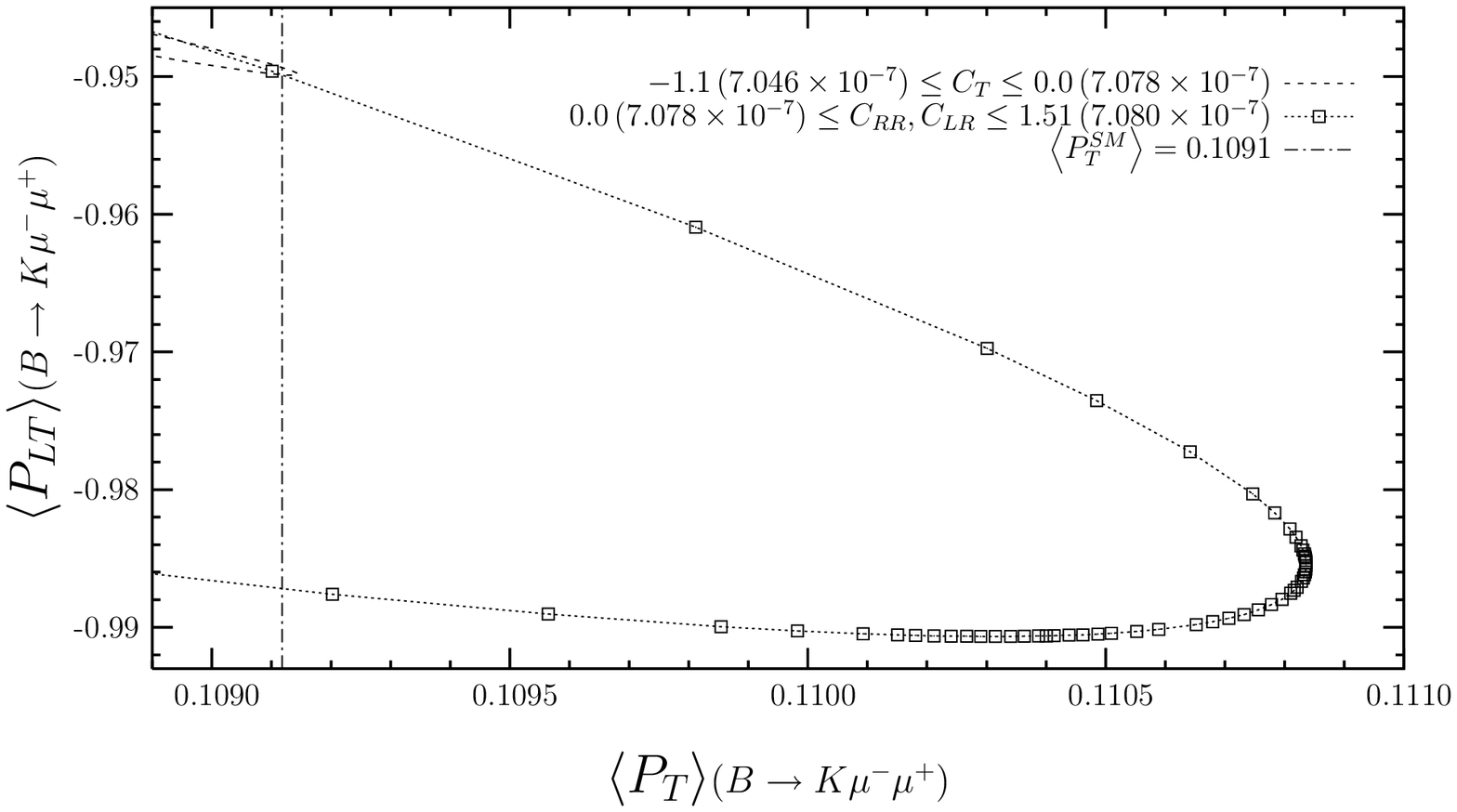}
\vskip 7.8 cm
\caption{}
\end{figure}

\begin{figure}
\vskip 1.5 cm
    \includegraphics{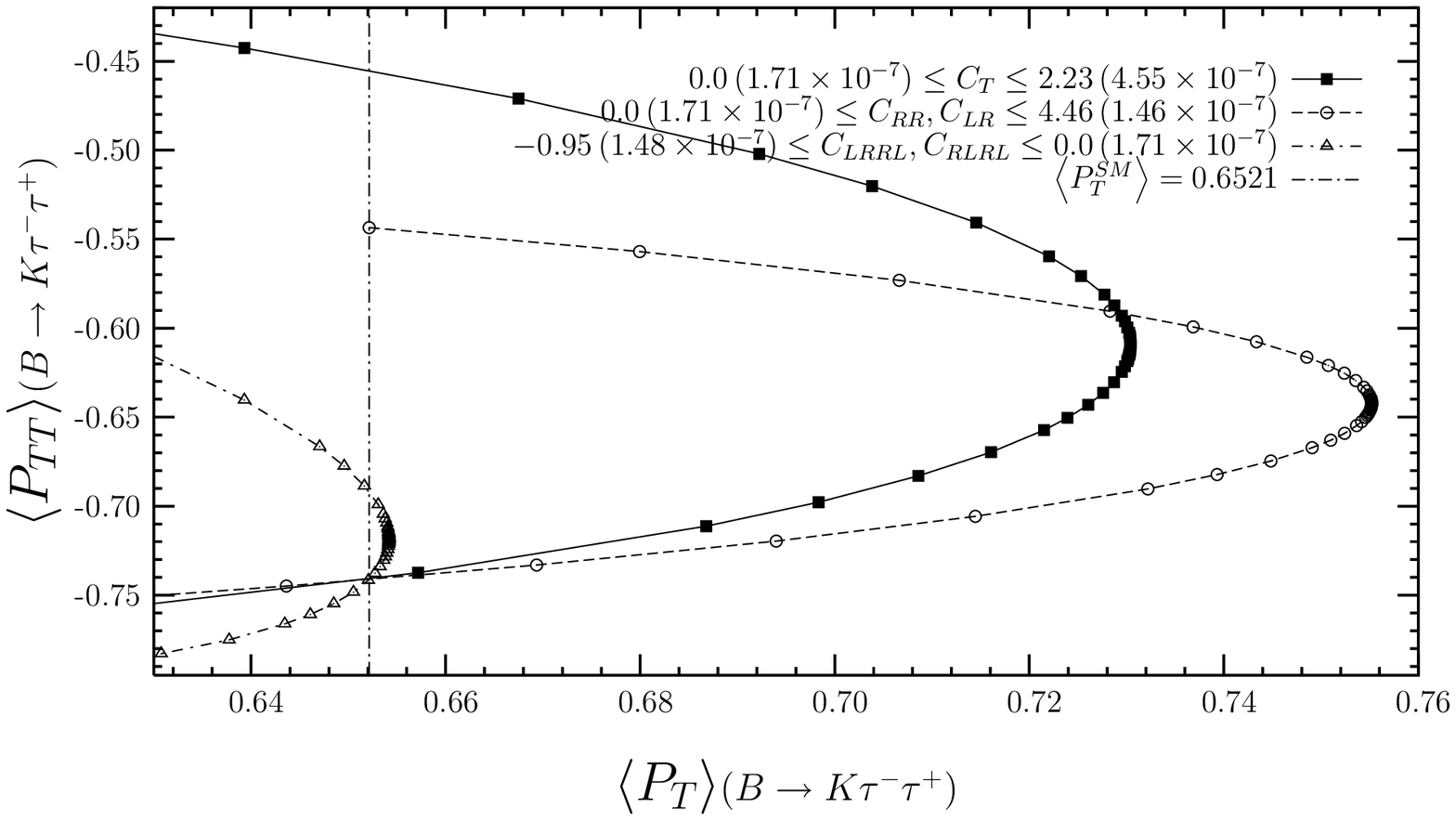}
\vskip 7.8cm
\caption{}
\end{figure}

\begin{figure}
\vskip 2.5 cm
    \includegraphics{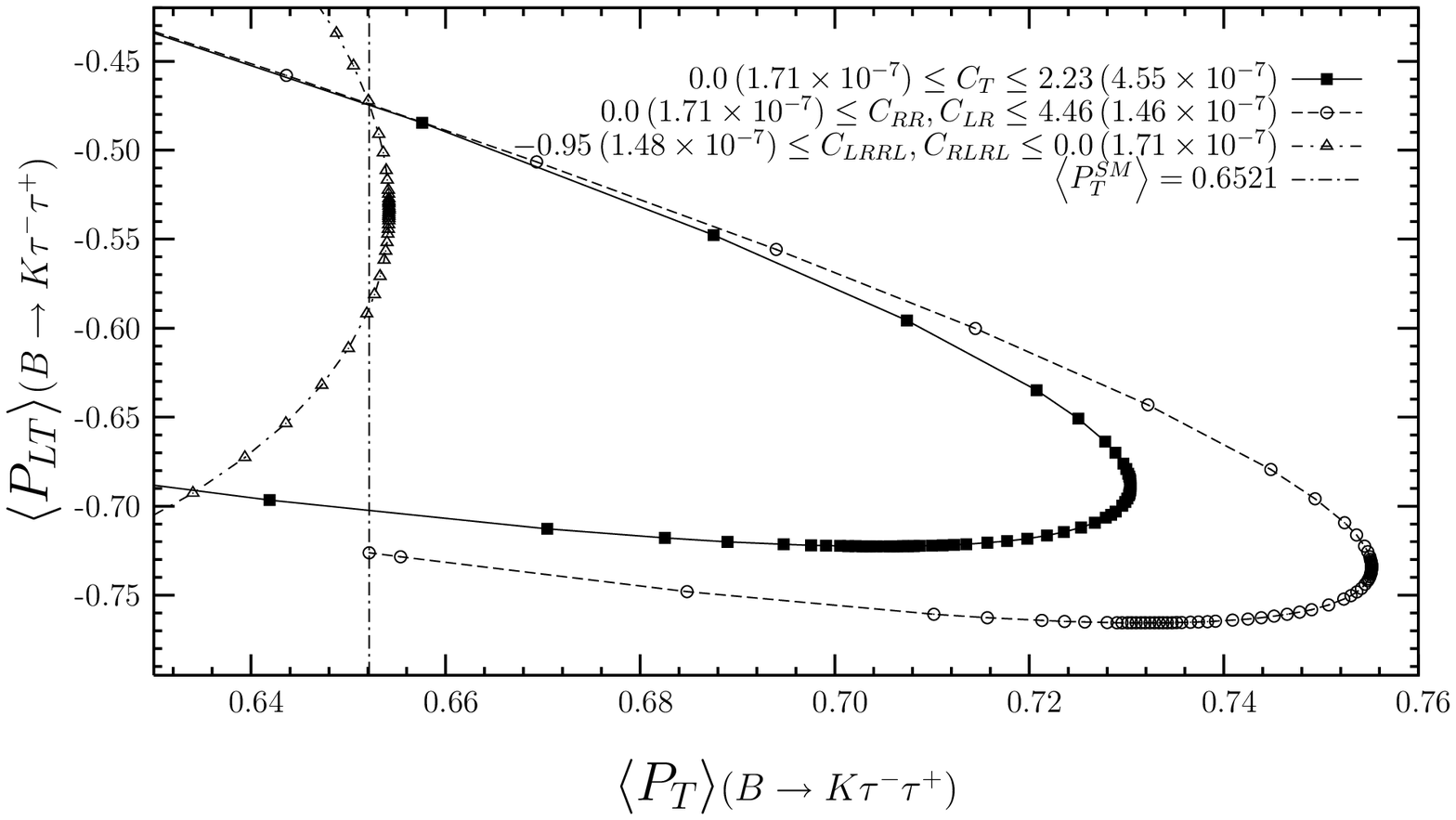}
\vskip 7.8 cm
\caption{}
\end{figure}

\begin{figure}
\vskip 2.5 cm
    \includegraphics{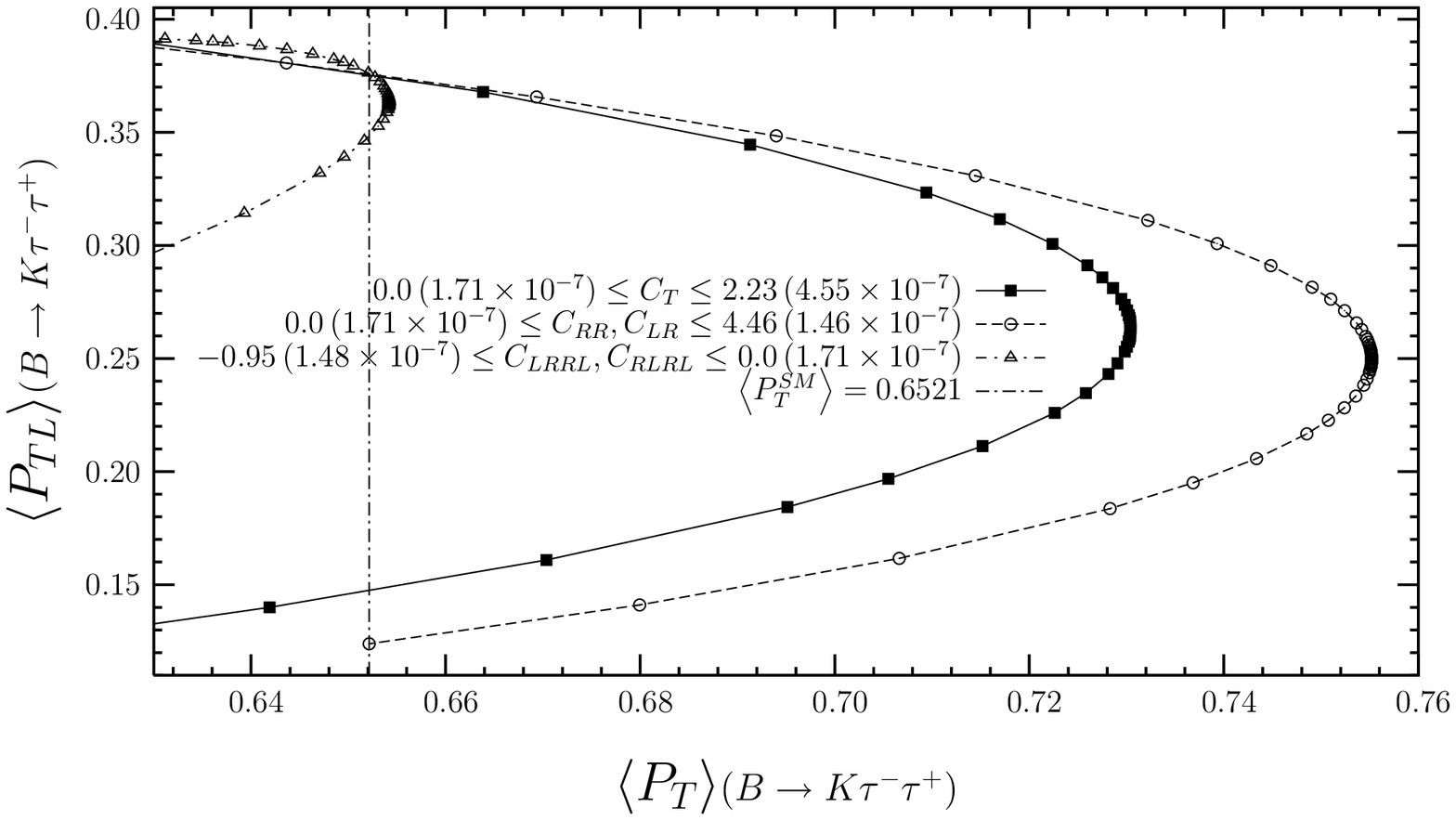}
\vskip 7.8 cm
\caption{}
\end{figure}

\end{document}